\documentclass[12pt]{mychapter}

\newcommand{\vdd}{V$_{DD}$\xspace}
\newcommand{\halfvdd}{$\frac{1}{2}$V$_{DD}$\xspace}
\newcommand{\halfvddpd}{$\frac{1}{2}$V$_{DD}+\delta$\xspace}

\newcommand{\src}{\texttt{src}\xspace}
\newcommand{\dst}{\texttt{dst}\xspace}
\newcommand{\tmp}{\texttt{tmp}\xspace}

\newcommand{\cmdact}{\texttt{ACTIVATE}\xspace}
\newcommand{\cmdpre}{\texttt{PRECHARGE}\xspace}
\newcommand{\cmdwr}{\texttt{WRITE}\xspace}
\newcommand{\cmdrd}{\texttt{READ}\xspace}
\newcommand{\cmdtr}{\texttt{TRANSFER}\xspace}

\newcommand{\mcpy}{\texttt{memcopy}\xspace}
\newcommand{\minit}{\texttt{meminit}\xspace}

\newcommand{\forkbench}{\texttt{forkbench}\xspace}
\newcommand{\ubsize}{$\mathcal{S}$\xspace}

\newcommand{\ttvdd}{$\frac{2}{3}$V$_{DD}$\xspace}
\newcommand{\otvdd}{$\frac{1}{3}$V$_{DD}$\xspace}

\newcommand{\band}{\textrm{\texttt{memand}}\xspace}
\newcommand{\bor}{\textrm{\texttt{memor}}\xspace}

\newcommand{\mrtwo}[1]{\multirow{2}{*}{#1}}
\newcommand{\mrthr}[1]{\multirow{3}{*}{#1}}

\newcommand{\fork}{\texttt{fork}\xspace}

\usepackage{tikz}
\usetikzlibrary{arrows}
\usetikzlibrary{patterns}
\usetikzlibrary{calc}
\usetikzlibrary{shapes}
\usetikzlibrary{positioning,fit}

\usepackage{circuitikz}

\tikzset{no padding/.style={inner sep=0pt, outer sep=0pt}}
\tikzset{rounded/.style={rounded corners=2pt}}
\tikzset{rounded1/.style={rounded corners=1pt}}

\tikzset{page/.style={draw,minimum width=1.75cm,minimum height=2.25cm, rounded}}

\tikzset{value/.style={draw,minimum width=2mm,minimum height=2mm, rounded1, fill=white}}

\title{\LARGE\sffamily \emph{The Processing Using Memory}
  Paradigm:\\In-DRAM Bulk Copy, Initialization, Bitwise AND and
  OR} \author{Vivek Seshadri --- Microsoft Research\\Onur Mutlu
  --- ETH Z{\"u}rich} \date{}
\begin{document}

\maketitle

\section*{Abstract}

In existing systems, the off-chip memory interface allows the
memory controller to perform only read or write
operations. Therefore, to perform any operation, the processor
must first read the source data and then write the result back to
memory after performing the operation. This approach consumes high
latency, bandwidth, and energy for operations that work on a large
amount of data. Several works have proposed techniques to process
data near memory by adding a small amount of compute logic closer
to the main memory chips. In this article, we describe two
techniques proposed by recent works that take this approach of
processing in memory further by exploiting the underlying
operation of the main memory technology to perform more complex
tasks. First, we describe RowClone, a mechanism that exploits DRAM
technology to perform bulk copy and initialization operations
completely inside main memory. We then describe a complementary work
that uses DRAM to perform bulk bitwise AND and OR operations
inside main memory. These two techniques significantly improve the
performance and energy efficiency of the respective operations.

\section{Introduction}

In modern systems, the channel that connects the processor and
off-chip main memory is a critical bottleneck for both performance
and energy-efficiency. First, the channel has limited data
bandwidth. Increasing this available bandwidth requires increasing
the number of channels or the width of each channel or the
frequency of the channel. All these approaches significantly
increase the cost of the system and are not scalable. Second, a
significant fraction of the energy consumed in performing an
operation is spent on moving data over the off-chip memory
channel~\cite{bill-dally}.

To address this problem, many prior and recent
works~\cite{pim-enabled-insts,pim-graph,top-pim,nda,msa3d,spmm-mul-lim,data-access-opt-pim,tom,hrl,gp-simd,ndp-architecture,pim-analytics,nda,jafar,data-reorg-3d-stack,smla,lim-computer,non-von-machine,iram,execube,active-pages,pim-terasys,cram,bitwise-cal,rowclone,pica,continuous-run-ahead,emc}
have proposed techniques to process data near memory, an approach
widely referred to as \emph{Processing in Memory} or PiM. The idea
behind PiM is to add a small amount of compute logic close to the
memory chips and use that logic to perform simple yet
bandwidth-intensive and/or latency-sensitive operations. The
premise is that being close to the memory chips, the PiM module
will have much higher bandwidth and lower latency to memory than
the regular processor. Consequently, PiM can 1)~perform
bandwidth-intensive and latency-sensitive operations faster and
2)~reduce the off-chip memory bandwidth requirements of such
operations.  As a result, PiM significantly improves both overall
system performance and energy efficiency.

In this article, we focus our attention on two works that push the
notion of processing in memory deeper by exploiting the underlying
operation of the main memory technology to perform more complex
tasks. We will refer to this approach as \emph{Processing using
  Memory} or PuM. Unlike PiM, which adds new logic structures near
memory to perform computation, the key idea behind PuM is to
exploit some of the peripheral structures \emph{already} existing
inside memory devices (with minimal changes) to perform other
tasks.

The first work that we will discuss in this article is
RowClone~\cite{rowclone}, a mechanism that exploits DRAM
technology to perform bulk data copy and initialization completely
inside DRAM. Such bulk copy and initialization operations are
triggered by many applications (e.g., bulk zeroing) and
system-level functions (e.g., page copy operations). Despite the
fact that these operations require no computation, existing system
must necessarily read and write the required data over the main
memory channel. In fact, even with a high-speed memory bus
(DDR4-2133) a simple 4~KB copy operation can take close to half a
micro second for just the data transfers on the memory channel. By
performing such operations completely inside main memory, RowClone
eliminates the need for any data transfer on the memory channel,
thereby significantly improving performance and energy-efficiency.

The second work that we will discuss in this article is a
mechanism to perform bulk bitwise AND and OR operations completely
inside DRAM~\cite{bitwise-cal}. Bitwise operations are an
important component of modern day programming. Many applications
(e.g., bitmap indices) rely on bitwise operations on large
bitvectors to achieve high performance. Similar to bulk copy or
initialization, the throughput of bulk bitwise operations in
existing systems is also limited by the available memory
bandwidth. The \emph{In-DRAM AND-OR} mechanism (IDAO) avoids
the need to transfer large amounts of data on the memory channel
to perform these operations. Similar to RowClone, IDAO enables an
order of magnitude improvement in the performance of bulk bitwise
operations. We will describe these two works in detail in this
article.

In this article, we will discuss the following things.
\begin{itemize}
\item We motivate the need for reducing data movement and how
  processing near memory helps in achieving that goal
  (Section~\ref{sec:pim}). We will briefly describe a set of
  recent works that have pushed the idea of processing near memory
  deeper by using the underlying memory technologies (e.g., DRAM,
  STT-MRAM, PCM) to perform tasks more complex than just storing
  data (Section~\ref{sec:pum}).
\item As the major focus of this article is on the PuM works that
  build on DRAM, we provide a brief background on modern DRAM
  organization and operation that is sufficient to understand the
  mechanisms (Section~\ref{sec:background}).
\item We describe the two mechanisms, RowClone (in-DRAM bulk copy
  and initialization) and In-DRAM-AND-OR (in-DRAM bulk bitwise AND
  and OR) in detail in Sections~\ref{sec:rowclone} and
  \ref{sec:and-or}, respectively.
\item We describe a number of applications for the two mechanisms
  and quantitative evaluations showing that they improve
  performance and energy-efficiency compared to existing systems.
\end{itemize}

\section{Processing in Memory}
\label{sec:pim}

Data movement contributes a major fraction of the execution time
and energy consumption of many programs. The farther the data is
from the processing engine (e.g., CPU), the more the contribution
of data movement towards execution time and energy
consumption. While most programs aim to keep their active working
set as close to the processing engine as possible (say the L1
cache), for applications with working sets larger than the on-chip
cache size, the data typically resides in main memory.

Unfortunately, main memory latency is not scaling commensurately
with the remaining resources in the system, namely, the compute
power and memory capacity. As a result, the performance of most
large-working-set applications is limited by main memory latency
and/or bandwidth. For instance, just transferring a single page
(4~KB) of data from DRAM can consume between a quarter and half a
microsecond even with high speed memory interfaces
(DDR4-2133~\cite{ddr4}). During this time, the processor can
potentially execute hundreds to thousands of instructions. With
respect to energy, while performing a 64-bit double precision
floating point operation typically consumes few tens of pico
joules, accessing 64-bits of data from off-chip DRAM consumes few
tens of nano joules (3 orders of magnitude more
energy)~\cite{bill-dally}.

One of the solutions to address this problem is to add support to
process data closer to memory, especially for operations that
access large amounts of data. This approach is generally referred
to as \emph{Processing in Memory (PiM)} or \emph{Near Data
  Processing}. The high-level idea behind PiM is to add a small
piece of compute logic closer to memory that has much higher
bandwidth to memory than the main processor. Prior research has
proposed two broad ways of implementing PiM: 1)~Integrating
processing logic into the memory chips, and 2)~using 3D-stacked memory
architectures.

\subsection{Integrating Processing Logic in Memory}

Many works (e.g., Logic-in-Memory Computer~\cite{lim-computer},
NON-VON Database Machine~\cite{non-von-machine},
EXECUBE~\cite{execube}, Terasys~\cite{pim-terasys}, Intelligent
RAM~\cite{iram}, Active Pages~\cite{active-pages},
FlexRAM~\cite{flexram,programming-flexram}, Computational
RAM~\cite{cram}, DIVA~\cite{diva} ) have proposed mechanisms and
models to add processing logic close to memory. The idea is to
integrate memory and CPU on the same chip by designing the CPU
using the memory process technology. The reduced data movement
allows these approaches to enable low-latency, high-bandwidth, and
low-energy data communication. However, they suffer from two key
shortcomings.

First, this approach of integrating processor on the same chip as
memory greatly increases the overall cost of the system. Second,
DRAM vendors use a high-density process to minimize
cost-per-bit. Unfortunately, high-density DRAM process is not
suitable for building high-speed logic~\cite{iram}. As a result,
this approach is not suitable for building a general purpose
processor near memory, at least with modern logic and high-density
DRAM technologies.

\subsection{3D-Stacked DRAM Architectures}

Some recent DRAM architectures~\cite{3d-stacking,hmc,hbm,smla} use
3D-stacking technology to stack multiple DRAM chips on top of the
processor chip or a separate logic layer. These architectures
offer much higher bandwidth to the logic layer compared to
traditional off-chip interfaces. This enables an opportunity to
offload some computation to the logic layer, thereby improving
performance. In fact, many recent works have proposed mechanisms
to improve and exploit such architectures
(e.g.,~\cite{pim-enabled-insts,pim-graph,top-pim,nda,msa3d,spmm-mul-lim,data-access-opt-pim,tom,hrl,gp-simd,ndp-architecture,pim-analytics,nda,jafar,data-reorg-3d-stack,smla,gpu-pim,lazypim,pica}).
3D-stacking enables much higher bandwidth between the logic layer
and the memory chips, compared to traditional
architectures. However, 3D-stacked architectures still require
data to be transferred outside the DRAM chip, and hence can be
bandwidth-limited. In addition, thermal factors constrain the
number of chips that can be stacked, thereby limiting the memory
capacity. As a result, multiple 3D-stacked DRAMs are required to
scale to large workloads. Despite these limitations, this approach
seems to be the most viable way of implementing processing in
memory in modern systems.

\section{Processing Using Memory}
\label{sec:pum}

In this article, we introduce a new class of work that pushes the
idea of PiM further by exploiting the underlying memory operation
to perform more complex operations than just data storage. We
refer to this class of works as \emph{Processing using Memory
  (PuM)}.

Reducing cost-per-bit is a first order design constraint for most
memory technologies. As a result, the memory cells are
small. Therefore, most memory devices use significant amount of
sensing and peripheral logic to extract data from the memory
cells. The key idea behind PuM is to use these logic structures
and their operation to perform some additional tasks.

It is clear that unlike PiM, which can potentially be designed to
perform any task, PuM can only enable some limited
functionality. However, for tasks that can be performed by PuM,
PuM has two advantages over PiM. First, as PuM exploits the
underlying operation of memory, it incurs much lower cost than
PiM. Second, unlike PiM, PuM does not have to read any data out of
the memory chips. As a result, the PuM approach is possibly the
most energy efficient way of performing the respective operations.

Building on top of DRAM, which is the technology ubiquitously used
to build main memory in modern systems, two recent works take the
PuM approach to accelerate certain important primitives:
1)~RowClone~\cite{rowclone}, which performs bulk copy and
initialization operations completely inside DRAM, and
2)~IDAO~\cite{bitwise-cal}, which performs bulk bitwise AND/OR
operations completely inside DRAM. Both these works exploit the
operation of the DRAM sense amplifier and the internal
organization of DRAM to perform the respective operations. We will
discuss these two works in significant detail in this article.

Similar to these works, there are others that build on various
other memory technologies. Pinatubo~\cite{pinatubo} exploits phase
change memory
(PCM)~\cite{pcm1,pcm2,pcm3,pcm4,pcm5,pcm-ibm,pcm-scalable}
architecture to perform bitwise operations efficiently inside
PCM. Pinatubo enhances the PCM sense amplifiers to sense fine
grained differences in resistance and use this to perform bitwise
operations on multiple cells connected to the same sense
amplifier. As we will describe in this article, bitwise operations
are critical for many important data structures like bitmap
indices. Kang et al.~\cite{sramsod} propose a mechanism to exploit
SRAM architecture to accelerate the primitive ``sum of absolute
differences''. ISAAC~\cite{isaac} is a mechanism to accelerate
vector dot product operations using a memristor array. ISAAC uses
the crossbar structure of a memristor array and its analog
operation to efficiently perform dot products. These operations
are heavily used in many important applications including deep
neural networks.

In the subsequent sections, we will focus our attention on
RowClone and IDAO. We will first provide the necessary background
on DRAM design and then describe how these mechanisms
work.

\section{Background on DRAM}
\label{sec:background}

In this section, we  describe the necessary background to
modern DRAM architecture and its implementation. While we focus
our attention primarily on commodity DRAM design (i.e., the DDRx
interface), most DRAM architectures use very similar design
approaches and vary only in higher-level design choices. As a
result, the mechanisms we describe in the subsequent sections can
be extended to any DRAM architecture. There has been significant
recent research in DRAM architectures and the interested reader
can find details about various aspects of DRAM in multiple recent
publications~\cite{salp,tl-dram,al-dram,gsdram,dsarp,ramulator,data-retention,parbor,fly-dram,efficacy-error-techniques,raidr,chargecache,avatar}.

\subsection{High-level Organization of the Memory System}

\begin{figure}[h]
  \centering
  \includegraphics{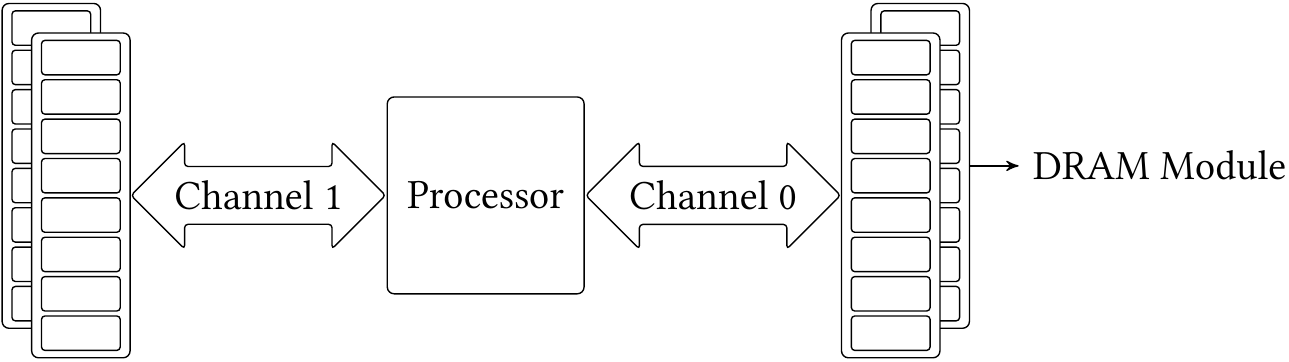}
  \caption{High-level organization of the memory subsystem}
  \label{fig:high-level-mem-org}
\end{figure}

Figure~\ref{fig:high-level-mem-org} shows the organization of the
memory subsystem in a modern system. At a high level, each
processor chip consists of one of more off-chip memory
\emph{channels}. Each memory channel consists of its own set of
\emph{command}, \emph{address}, and \emph{data} buses. Depending
on the design of the processor, there can be either an independent
memory controller for each memory channel or a single memory
controller for all memory channels. All modules connected to a
channel share the buses of the channel. Each module consists of
many DRAM devices (or chips). Most of this section is dedicated to
describing the design of a modern DRAM chip. In
Section~\ref{sec:dram-module}, we present more details of the
module organization of commodity DRAM.

\subsection{DRAM Chip}
\label{sec:dram-chip}

A modern DRAM chip consists of a hierarchy of structures: DRAM
\emph{cells}, \emph{tiles/MATs}, \emph{subarrays}, and
\emph{banks}. In this section, we  describe the design of a
modern DRAM chip in a bottom-up fashion, starting from a single
DRAM cell and its operation.

\subsubsection{DRAM Cell and Sense Amplifier}

\begin{figure}[h]
  \centering
  \includegraphics{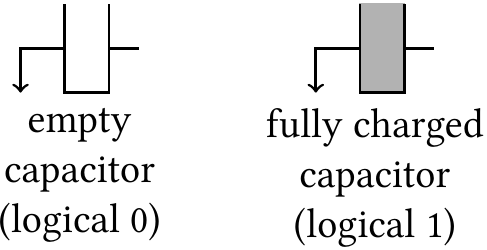}
  \caption[Capacitor]{Two states of a DRAM cell}
  \label{fig:cell-states}
\end{figure}

At the lowest level, DRAM technology uses capacitors to store
information. Specifically, it uses the two extreme states of a
capacitor, namely, the \emph{empty} and the \emph{fully charged}
states to store a single bit of information. For instance, an
empty capacitor can denote a logical value of 0, and a fully
charged capacitor can denote a logical value of 1.
Figure~\ref{fig:cell-states} shows the two extreme states of a
capacitor.

Unfortunately, the capacitors used for DRAM chips are small, and
will get smaller with each new generation. As a result, the amount
of charge that can be stored in the capacitor, and hence the
difference between the two states is also very small. In addition,
the capacitor can potentially lose its state after it is
accessed. Therefore, to extract the state of the capacitor, DRAM
manufacturers use a component called \emph{sense amplifier}.

\begin{figure}[h]
  \centering
  \begin{minipage}{5cm}
    \centering
    \includegraphics{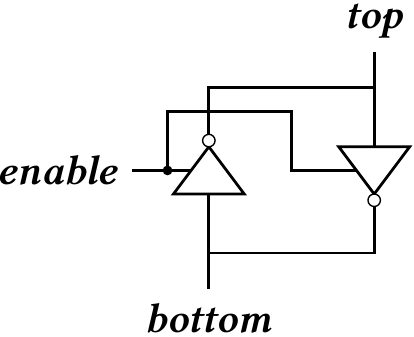}
    \caption{Sense amplifier}
    \label{fig:sense-amp}
  \end{minipage}\quad
  \begin{minipage}{9cm}
    \centering
    \includegraphics{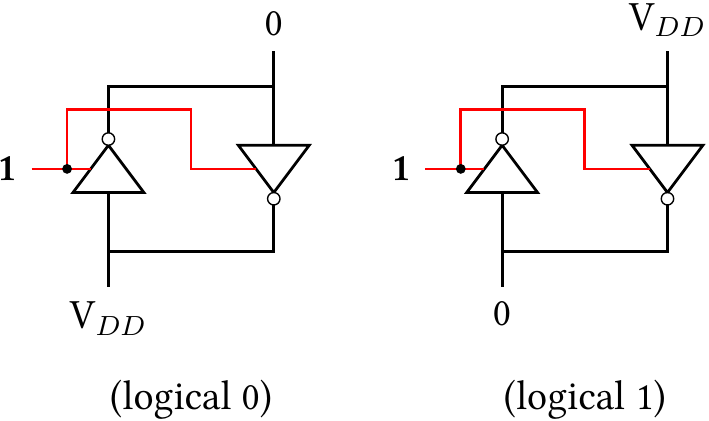}
    \caption{Stable states of a sense amplifier}
    \label{fig:sense-amp-states}
  \end{minipage}
\end{figure}

Figure~\ref{fig:sense-amp} shows a sense amplifier. A sense
amplifier contains two inverters which are connected together such
that the output of one inverter is connected to the input of the
other and vice versa. The sense amplifier also has an enable
signal that determines if the inverters are active. When enabled,
the sense amplifier has two stable states, as shown in
Figure~\ref{fig:sense-amp-states}. In both these stable states,
each inverter takes a logical value and feeds the other inverter
with the negated input.

\begin{figure}[h]
  \centering
  \includegraphics{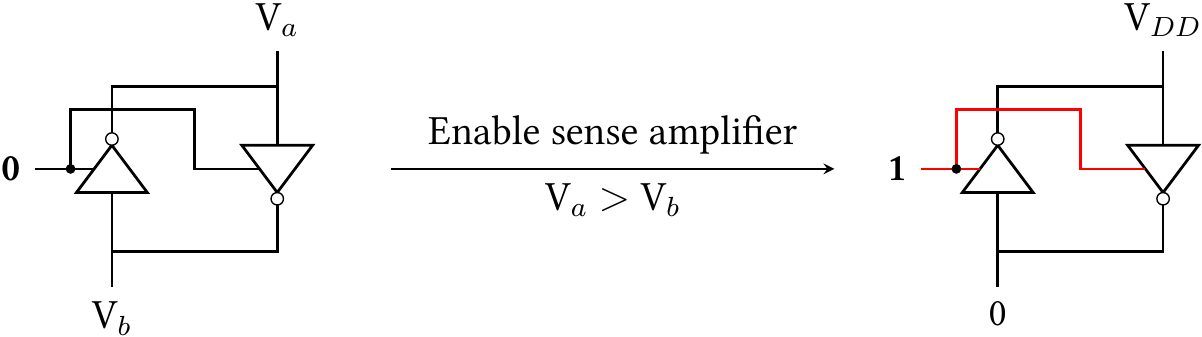}
  \caption{Operation of the sense amplifier}
  \label{fig:sense-amp-operation}
\end{figure}

Figure~\ref{fig:sense-amp-operation} shows the operation of the
sense amplifier from a disabled state. In the initial disabled
state, we assume that the voltage level of the top terminal
(V$_a$) is higher than that of the bottom terminal (V$_b$).
When the sense amplifier is enabled in this state, it
\emph{senses} the difference between the two terminals and
\emph{amplifies} the difference until it reaches one of the stable
states (hence the name ``sense amplifier'').  

\subsubsection{DRAM Cell Operation: The \texttt{ACTIVATE-PRECHARGE}
  cycle}
\label{sec:cell-operation}

\begin{figure}[h]
  \centering
  \includegraphics{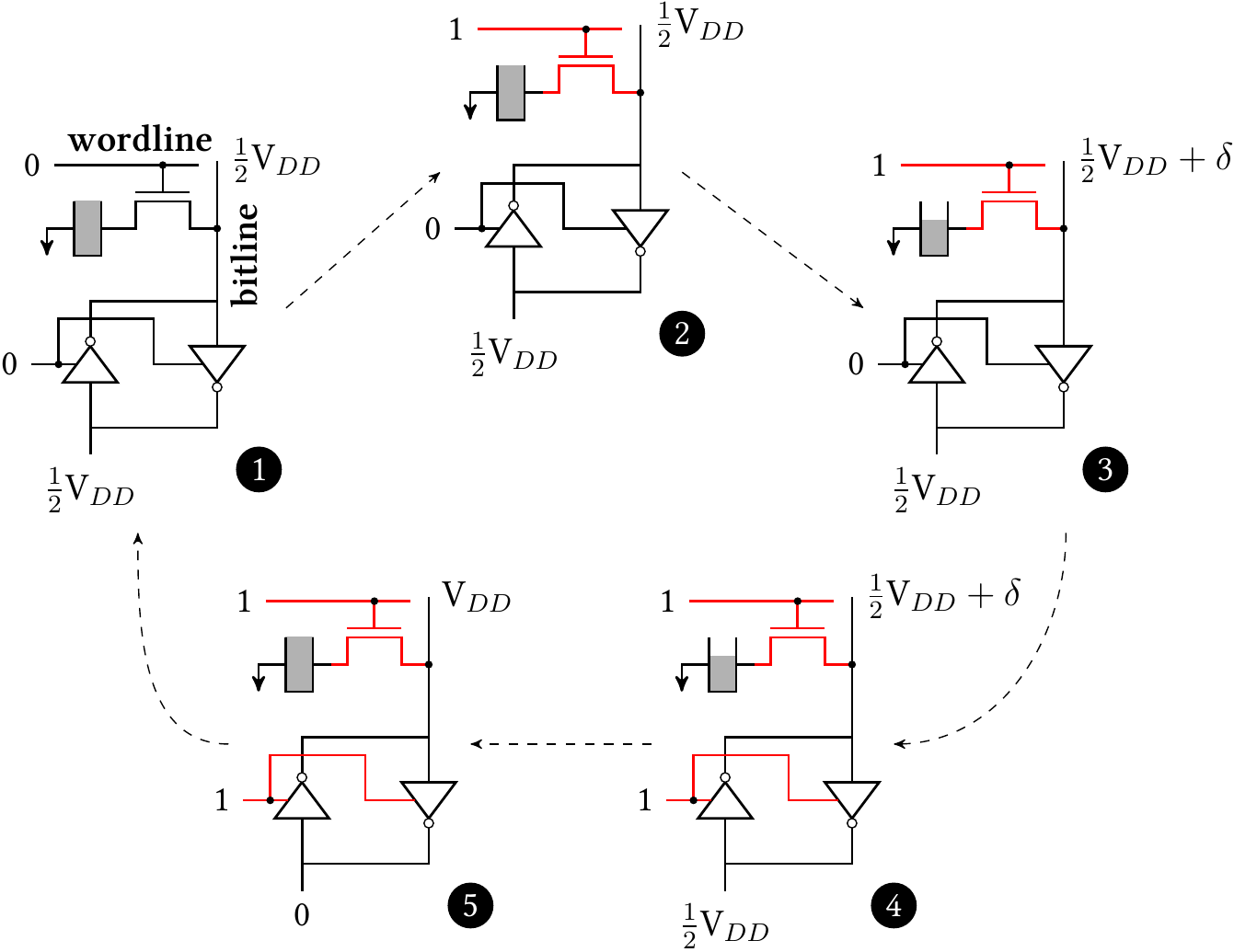}
  \caption{Operation of a DRAM cell and sense amplifier}
  \label{fig:cell-operation}
\end{figure}

DRAM technology uses a simple mechanism that converts the logical
state of a capacitor into a logical state of the sense
amplifier. Data can then be accessed from the sense amplifier
(since it is in a stable state). Figure~\ref{fig:cell-operation}
shows the connection between a DRAM cell and the sense amplifier
and the sequence of states involved in converting the cell state
to the sense amplifier state.

As shown in the figure (state \ding{202}), the capacitor is
connected to an access transistor that acts as a switch between
the capacitor and the sense amplifier. The transistor is
controller by a wire called \emph{wordline}. The wire that
connects the transistor to the top end of the sense amplifier is
called \emph{bitline}. In the initial state \ding{202}, the
wordline is lowered, the sense amplifier is disabled and both ends
of the sense amplifier are maintained at a voltage level of
\halfvdd. We assume that the capacitor is initially fully charged
(the operation is similar if the capacitor was empty). This state
is referred to as the \emph{precharged} state. An access to the
cell is triggered by a command called \cmdact. Upon receiving an
\cmdact, the corresponding wordline is first raised (state
\ding{203}). This connects the capacitor to the bitline. In the
ensuing phase called \emph{charge sharing} (state \ding{204}),
charge flows from the capacitor to the bitline, raising the
voltage level on the bitline (top end of the sense amplifier) to
\halfvddpd. After charge sharing, the sense amplifier is enabled
(state \ding{205}). The sense amplifier detects the difference in
voltage levels between its two ends and amplifies the deviation,
till it reaches the stable state where the top end is at \vdd
(state \ding{206}). Since the capacitor is still connected to the
bitline, the charge on the capacitor is also fully restored. We
shortly describe how the data can be accessed from the sense
amplifier. However, once the access to the cell is complete, the
cell is taken back to the original precharged state using the
command called \cmdpre. Upon receiving a \cmdpre, the wordline is
first lowered, thereby disconnecting the cell from the sense
amplifier. Then, the two ends of the sense amplifier are driven to
\halfvdd using a precharge unit (not shown in the figure for
brevity).

\subsubsection{DRAM MAT/Tile: The Open Bitline Architecture}
\label{sec:dram-mat}

A major goal of DRAM manufacturers is to maximize the density of
the DRAM chips while adhering to certain latency constraints
(described in Section~\ref{sec:dram-timing-constraints}). There
are two costly components in the setup described in the previous
section. The first component is the sense amplifier itself. Each
sense amplifier is around two orders of magnitude larger than a
single DRAM cell~\cite{rambus-power,tl-dram}. Second, the state of the
wordline is a function of the address that is currently being
accessed. The logic that is necessary to implement this function
(for each cell) is expensive.

\begin{figure}[h]
  \centering
  \includegraphics{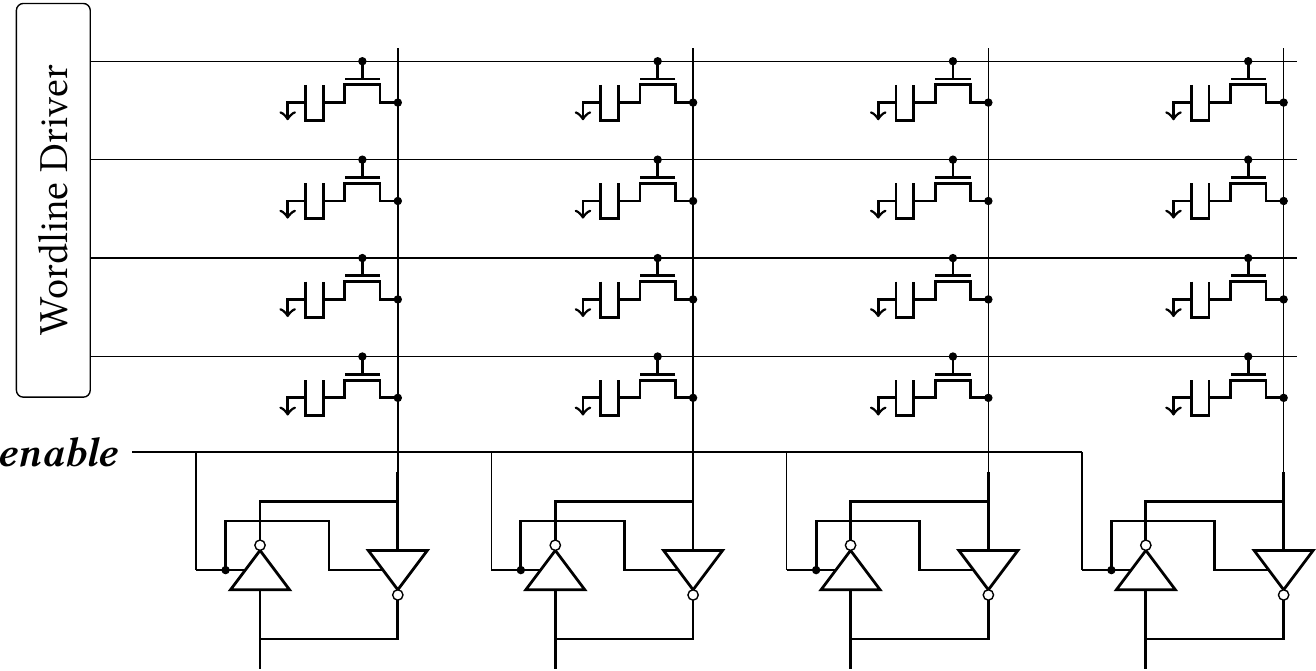}
  \caption{A 2-D array of DRAM cells}
  \label{fig:dram-mat}
\end{figure}

In order to reduce the overall cost of these two components, they
are shared by many DRAM cells. Specifically, each sense amplifier
is shared by a column of DRAM cells. In other words, all the cells in
a single column are connected to the same bitline. Similarly, each
wordline is shared by a row of DRAM cells. Together, this
organization consists of a 2-D array of DRAM cells connected to a
row of sense amplifiers and a column of wordline
drivers. Figure~\ref{fig:dram-mat} shows this organization with a
$4 \times 4$ 2-D array.

\begin{figure}[h]
  \centering
  \includegraphics{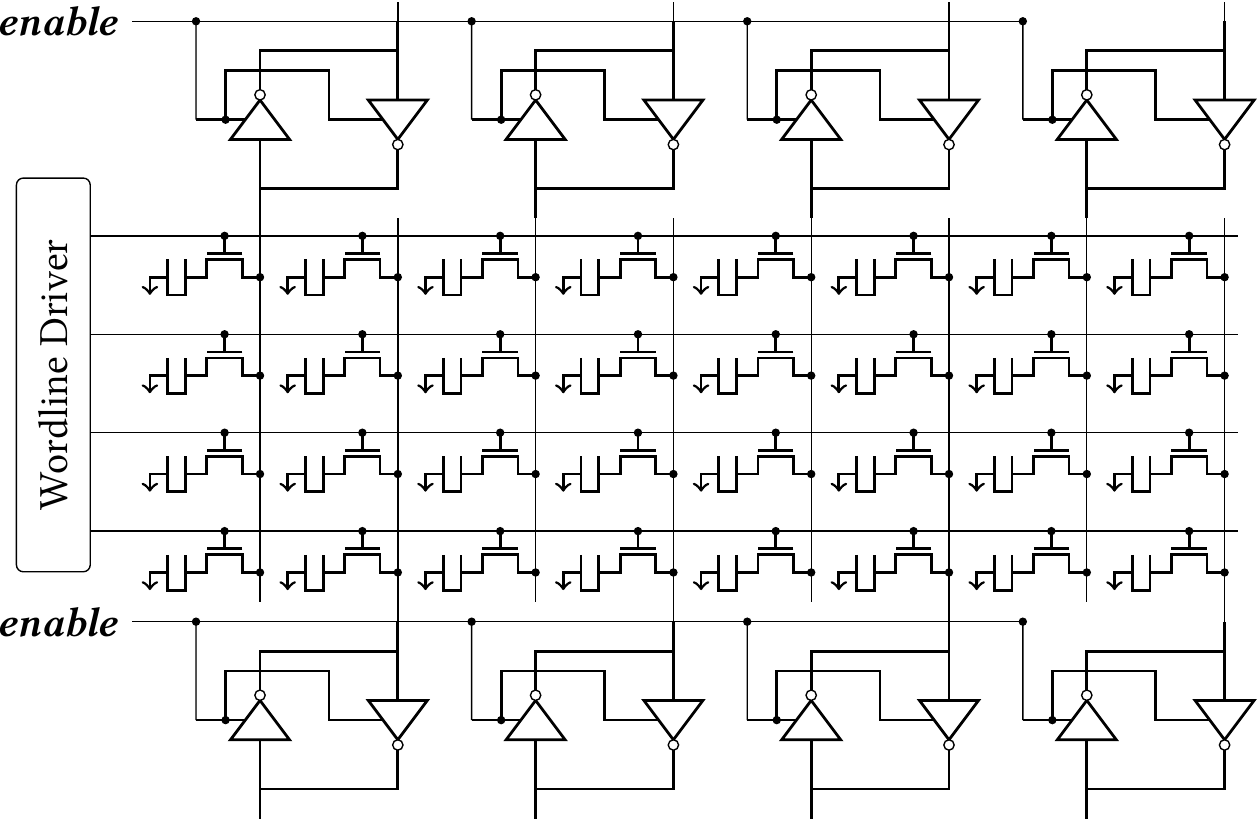}
  \caption{A DRAM MAT/Tile: The open bitline architecture}
  \label{fig:dram-mat-oba}
\end{figure}

To further reduce the overall cost of the sense amplifiers and the
wordline driver, modern DRAM chips use an architecture called the
\emph{open bitline architecture}. This architecture exploits two
observations. First, the sense amplifier is wider than the DRAM
cells. This difference in width results in a white space near each
column of cells. Second, the sense amplifier is
symmetric. Therefore, cells can also be connected to the bottom
part of the sense amplifier. Putting together these two
observations, we can pack twice as many cells in the same area
using the open bitline architecture, as shown in
Figure~\ref{fig:dram-mat-oba};

As shown in the figure, a 2-D array of DRAM cells is connected to
two rows of sense amplifiers: one on the top and one on the bottom
of the array. While all the cells in a given row share a common
wordline, half the cells in each row are connected to the top row
of sense amplifiers and the remaining half of the cells are
connected to the bottom row of sense amplifiers. This tightly
packed structure is called a DRAM
MAT/Tile~\cite{rethinking-dram,half-dram,salp}. In a modern DRAM
chip, each MAT typically is a $512 \times 512$ or $1024 \times
1024$ array. Multiple MATs are grouped together to form a larger
structure called a \emph{DRAM bank}, which we describe next.

\subsubsection{DRAM Bank}

\begin{figure}[h]
  \centering
  \includegraphics{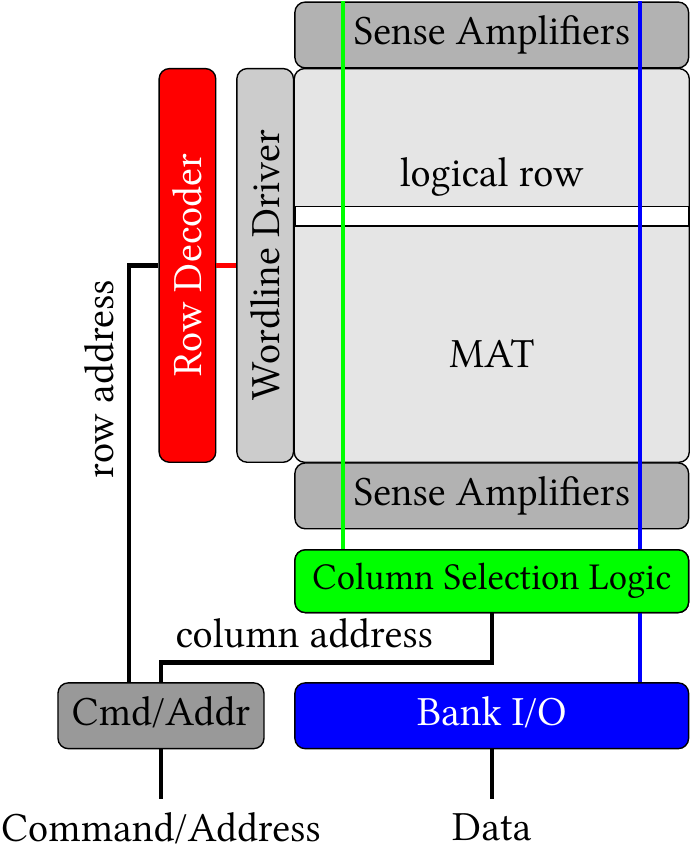}
  \caption{DRAM Bank: Logical view}
  \label{fig:dram-bank-logical}
\end{figure}

In most modern commodity DRAM interfaces~\cite{ddr3,ddr4}, a DRAM
bank is the smallest structure visible to the memory
controller. All commands related to data access are directed to a
specific bank. Logically, each DRAM bank is a large monolithic
structure with a 2-D array of DRAM cells connected to a single set
of sense amplifiers (also referred to as a row buffer). For
example, in a 2Gb DRAM chip with 8 banks, each bank has $2^{15}$
rows and each logical row has 8192 DRAM
cells. Figure~\ref{fig:dram-bank-logical} shows this logical view
of a bank.

In addition to the MAT, the array of sense amplifiers, and the
wordline driver, each bank also consists of some peripheral
structures to decode DRAM commands and addresses, and manage the
input/output to the DRAM bank. Specifically, each bank has a
\emph{row decoder} to decode the row address of row-level commands
(e.g., \cmdact). Each data access command (\cmdrd and \cmdwr)
accesses only a part of a DRAM row. Such individual parts are
referred to as \emph{columns}. With each data access command, the
address of the column to be accessed is provided. This address is
decoded by the \emph{column selection logic}. Depending on which
column is selected, the corresponding piece of data is
communicated between the sense amplifiers and the bank I/O
logic. The bank I/O logic in turn acts as an interface between the
DRAM bank and the chip-level I/O logic.

\begin{figure}[h!]
  \centering
  \includegraphics{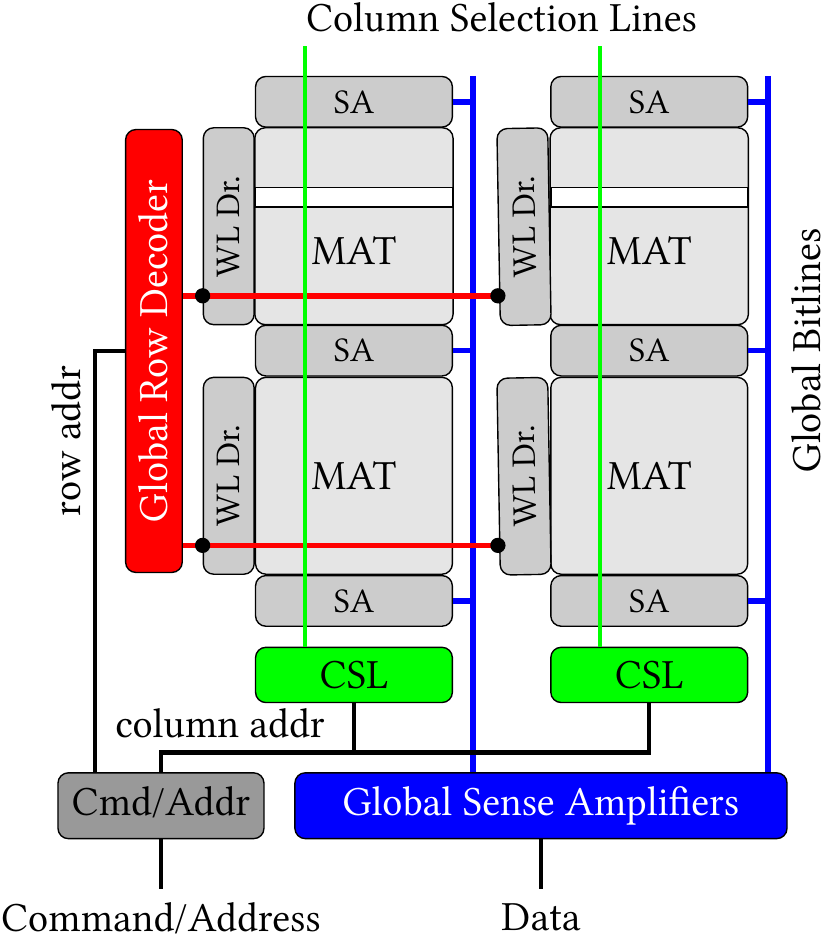}
  \caption{DRAM Bank: Physical implementation. In a real chip, the
    global bitlines run on top of the MATs in a separate metal
    layer. (components in figure are not to scale)}
  \label{fig:dram-bank-physical}
\end{figure}

Although the bank can logically be viewed as a single MAT,
building a single MAT of a very large dimension is practically not
feasible as it will require very long bitlines and
wordlines. Therefore, each bank is physically implemented as a 2-D
array of DRAM MATs. Figure~\ref{fig:dram-bank-physical} shows a
physical implementation of the DRAM bank with 4 MATs arranged in
$2 \times 2$ array. As shown in the figure, the output of the
global row decoder is sent to each row of MATs.  The bank I/O
logic, also known as the \emph{global sense amplifiers}, are
connected to all the MATs through a set of \emph{global
  bitlines}. As shown in the figure, each vertical collection of
MATs consists of its own columns selection logic and global
bitlines. In a real DRAM chip, the global bitlines run on top of
the MATs in a separate metal layer. One implication of this
division is that the data accessed by any command is split equally
across all the MATs in a single row of MATs.

\begin{figure}[h]
  \centering
  \includegraphics{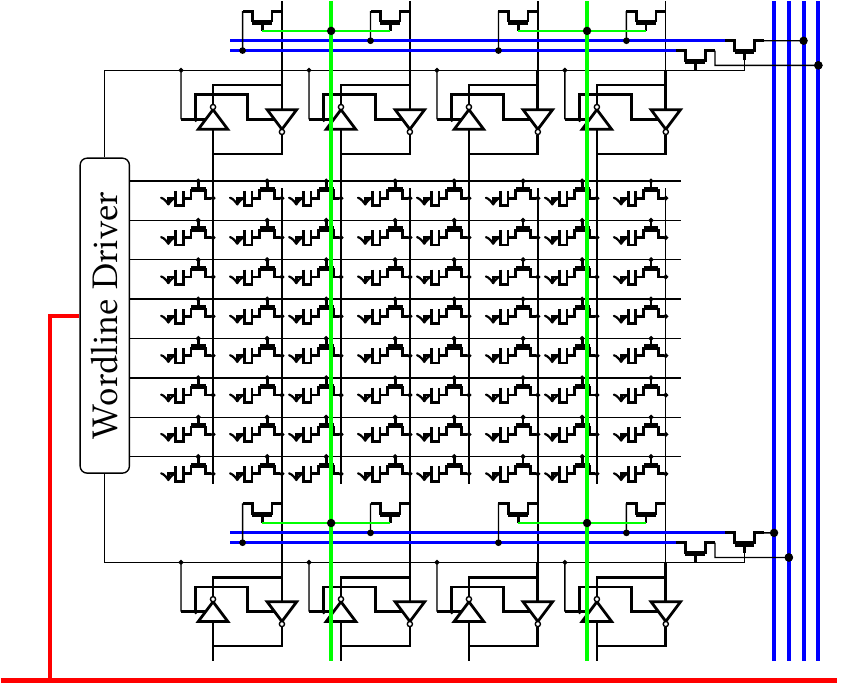}
  \caption{Detailed view of a MAT}
  \label{fig:dram-mat-zoomed}
\end{figure}

\begin{figure}[h]
  \centering
  \includegraphics{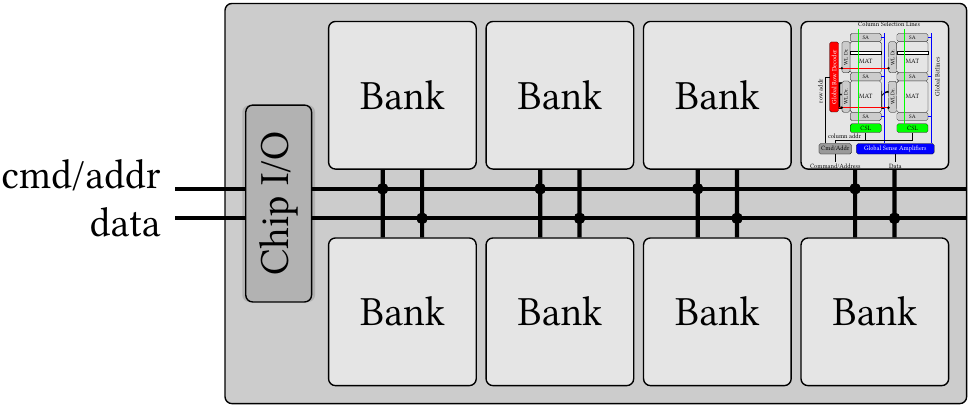}
  \caption{DRAM Chip}
  \label{fig:dram-chip}
\end{figure}

Figure~\ref{fig:dram-mat-zoomed} shows the zoomed-in version of a
DRAM MAT with the surrounding peripheral logic. Specifically, the
figure shows how each column selection line selects specific sense
amplifiers from a MAT and connects them to the global bitlines. It
should be noted that the width of the global bitlines for each MAT
(typically 8/16) is much smaller than that of the width of the MAT
(typically 512/1024). This is because the global bitlines span a
much longer distance and hence have to be thicker to ensure
integrity.

Each DRAM chip consist of multiple banks as shown in
Figure~\ref{fig:dram-chip}. All the banks share the chip's
internal command, address, and data buses. As mentioned before,
each bank operates mostly independently (except for operations that
involve the shared buses). The chip I/O manages the transfer of
data to and from the chip's internal bus to the memory
channel. The width of the chip output (typically 8 bits) is much
smaller than the output width of each bank (typically 64
bits). Any piece of data accessed from a DRAM bank is first
buffered at the chip I/O and sent out on the memory bus 8 bits at
a time. With the DDR (double data rate) technology, 8 bits are
sent out each half cycle. Therefore, it takes 4 cycles to transfer
64 bits of data from a DRAM chip I/O on to the memory channel.

\subsubsection{DRAM Commands: Accessing Data from a DRAM Chip}

To access a piece of data from a DRAM chip, the memory controller
must first identify the location of the data: the bank ID ($B$),
the row address ($R$) within the bank, and the column address
($C$) within the row. After identifying these pieces of
information, accessing the data involves three steps.

The first step is to issue a \cmdpre to the bank $B$. This step
prepares the bank for a data access by ensuring that all the sense
amplifiers are in the \emph{precharged} state
(Figure~\ref{fig:cell-operation}, state~\ding{202}). No wordline
within the bank is raised in this state.

The second step is to activate the row $R$ that contains the
data. This step is triggered by issuing a \cmdact to bank $B$ with
row address $R$. Upon receiving this command, the corresponding
bank feeds its global row decoder with the input $R$. The global
row decoder logic then raises the wordline of the DRAM row
corresponding to the address $R$ and enables the sense amplifiers
connected to that row. This triggers the DRAM cell operation
described in Section~\ref{sec:cell-operation}. At the end of the
activate operation the data from the entire row of DRAM cells is
copied to the corresponding array of sense amplifiers.

Finally, the third step is to access the data from the required
column. This is done by issuing a \cmdrd or \cmdwr command to the
bank with the column address $C$. Upon receiving a \cmdrd or
\cmdwr command, the corresponding address is fed to the column
selection logic. The column selection logic then raises the column
selection lines (Figure~\ref{fig:dram-mat-zoomed}) corresponding
to address $C$, thereby connecting those sense amplifiers to the
global sense amplifiers through the global bitlines. For a read
access, the global sense amplifiers sense the data from the MAT's
local sense amplifiers and transfer that data to the chip's
internal bus. For a write access, the global sense amplifiers read
the data from the chip's internal bus and force the MAT's local
sense amplifiers to the appropriate state.

Not all data accesses require all three steps. Specifically, if
the row to be accessed is already activated in the corresponding
bank, then the first two steps can be skipped and the data can be
directly accessed by issuing a \cmdrd or \cmdwr to the bank. For
this reason, the array of sense amplifiers are also referred to as
a \emph{row buffer}, and such an access that skips the first two
steps is called a \emph{row buffer hit}. Similarly, if the bank is
already in the precharged state, then the first step can be
skipped. Such an access is referred to as a \emph{row buffer
  miss}. Finally, if a different row is activated within the bank,
then all three steps have to be performed. Such an access is
referred to as a \emph{row buffer conflict}.

\subsubsection{DRAM Timing Constraints}
\label{sec:dram-timing-constraints}

Different operations within DRAM consume different amounts of
time. Therefore, after issuing a command, the memory controller
must wait for a sufficient amount of time before it can issue the
next command. Such wait times are managed by what are called the
\emph{timing constraints}. Timing constraints essentially dictate
the minimum amount of time between two commands issued to the same
bank/rank/channel. Table~\ref{table:timing-constraints} describes
some key timing constraints along with their values for the
DDR3-1600 interface.

\begin{table}[h]\small
  \centering
  \begin{tabular}{lrclp{2.2in}r}
  \toprule
  Name & \multicolumn{3}{c}{Constraint} & Description & Value (ns)\\
  \toprule
  \mrtwo{tRAS} & \mrtwo \cmdact & \mrtwo{$\rightarrow$} & \mrtwo \cmdpre & Time taken to complete a row
  activation operation in a bank & \mrtwo{35}\\
  \midrule
  \mrtwo{tRCD} & \mrtwo \cmdact & \mrtwo{$\rightarrow$} & \mrtwo {\cmdrd/\cmdwr} & Time between an activate
  command and column command to a bank & \mrtwo{15}\\
  \midrule
  \mrtwo{tRP} & \mrtwo \cmdpre & \mrtwo{$\rightarrow$} & \mrtwo \cmdact & Time taken to complete a precharge
  operation in a bank & \mrtwo{15}\\
  \midrule
  \mrthr{tWR} & \mrthr \cmdwr & \mrthr{$\rightarrow$} & \mrthr \cmdpre & Time taken to ensure that data is
  safely written to the DRAM cells after a write operation (\emph{write recovery}) & \mrthr{15}\\
  \bottomrule
\end{tabular}

  \caption[DDR3-1600 DRAM timing constraints]{Key DRAM timing constraints with their values for DDR3-1600}
  \label{table:timing-constraints}
\end{table}

\subsection{DRAM Module}
\label{sec:dram-module}

As mentioned before, each \cmdrd or \cmdwr command for a single
DRAM chip typically involves only 64 bits. In order to achieve
high memory bandwidth, commodity DRAM modules group several DRAM
chips (typically 4 or 8) together to form a \emph{rank} of DRAM
chips. The idea is to connect all chips of a single rank to the
same command and address buses, while providing each chip with an
independent data bus. In effect, all the chips within a rank
receive the same commands with same addresses, making the rank a
logically wide DRAM chip.

\begin{figure}[h]
  \centering
  \begin{tikzpicture}[semithick]

  \tikzset{chip/.style={draw,rounded corners=3pt,black!40,fill=black!20,
     minimum width=1.4cm, minimum height=1.75cm,outer sep=0pt,anchor=west}};
  \tikzset{wire/.style={thin,black!80}};
  
  \node (chip0) [chip] at (0,0) {};
  \foreach \x [count=\i] in {0,...,6} {
    \node (chip\i) [chip, xshift=2mm] at (chip\x.east) {};
  }

  \foreach \i in {0,...,7} {
    \coordinate (cmd\i) at ($(chip\i.south) + (-0.3,0)$);
    \coordinate (addr\i) at (chip\i.south);
    \coordinate (data\i) at ($(chip\i.south) + (0.3,0)$);
  }

  \coordinate (cmdorigin) at ($(chip0.south west) + (-1.5,-0.4)$);
  \coordinate (addrorigin) at ($(chip0.south west) + (-1.5,-1)$);
  \draw [wire] (cmdorigin) -- (cmdorigin-|chip7.south east);
  \draw [wire] (addrorigin) -- (addrorigin-|chip7.south east);

  \foreach \i in {0,...,7} {
    \coordinate (dataorg\i) at ($(chip0.south west) + (-1.5,-1.5 - 0.2*\i)$);
    \draw [wire] (dataorg\i) -- (dataorg\i-|chip7.south east);
  }

  \foreach \i in {0,...,7} {
    \node at (chip\i) {Chip \i};
    \draw [wire,fill] (cmd\i) -- (cmd\i|-cmdorigin) circle(1.5pt);
    \draw [wire,fill] (addr\i) -- (addr\i|-addrorigin) circle(1.5pt);
    \draw [wire,fill] (data\i) -- (data\i|-dataorg\i) circle(1.5pt);
  }

  \node at (cmdorigin) [xshift=3mm,fill=white,anchor=west] {\emph{cmd}};
  \node at (addrorigin) [xshift=3mm,fill=white,anchor=west] {\emph{addr}};
  \node at ($(dataorg1)!0.5!(dataorg6)$) [xshift=3mm,fill=white,anchor=west] {\emph{data} (64 bits)};

\end{tikzpicture}
  \caption{Organization of a DRAM rank}
  \label{fig:dram-rank}
\end{figure}
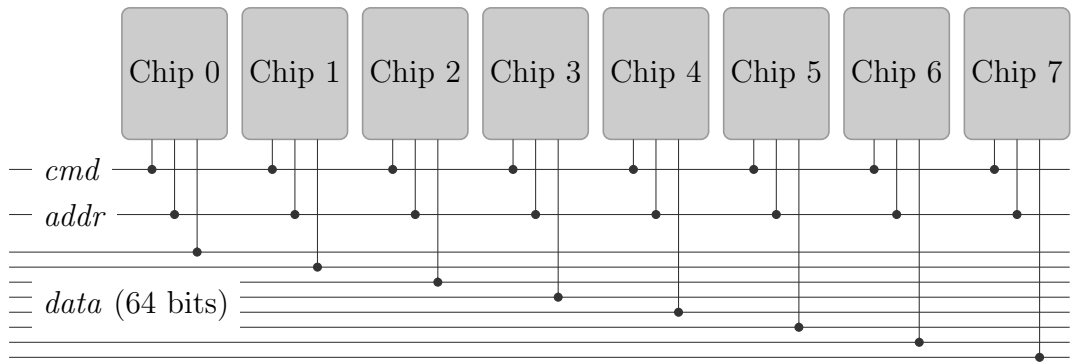

Figure~\ref{fig:dram-rank} shows the logical organization of a
DRAM rank. Most commodity DRAM ranks consist of 8
chips. Therefore, each \cmdrd or \cmdwr command accesses 64 bytes
of data, the typical cache line size in most processors.

\section{RowClone}
\label{sec:rowclone}

In this section, we present RowClone~\cite{rowclone}, a mechanism
that can perform bulk copy and initialization operations
completely inside DRAM. This approach obviates the need to
transfer large quantities of data on the memory channel, thereby
significantly improving the efficiency of a bulk copy
operation. As bulk data initialization (specifically bulk zeroing)
can be viewed as a special case of a bulk copy operation, RowClone
can be easily extended to perform such bulk initialization
operations with high efficiency.

RowClone consists of two independent mechanisms that exploit
several observations about DRAM organization and operation.  The
first mechanism, called the \emph{Fast Parallel Mode} (FPM),
efficiently copies data between two rows of DRAM cells that share
the same set of sense amplifiers (i.e., two rows within the same
subarray).  The second mechanism, called the \emph{Pipelined
  Serial Mode}, efficiently copies cache lines between two banks
within a module in a pipelined manner.  Although not as fast as
FPM, PSM has fewer constraints and hence is more generally
applicable. We now describe these two mechanisms in detail.

\subsection{Fast-Parallel Mode}
\label{sec:rowclone-fpm}

The Fast Parallel Mode (FPM) is based on the following three
observations about DRAM.
\begin{enumerate}
\item In a commodity DRAM module, each \cmdact command transfers
  data from a large number of DRAM cells (multiple kilo-bytes) to
  the corresponding array of sense amplifiers
  (Section~\ref{sec:dram-module}).
\item Several rows of DRAM cells share the same set of sense
  amplifiers (Section~\ref{sec:dram-mat}).
\item A DRAM cell is not strong enough to flip the state of the
  sense amplifier from one stable state to another stable
  state. In other words, if a cell is connected to an already
  activated sense amplifier (or bitline), then the data of the
  cell gets overwritten with the data on the sense amplifier.
\end{enumerate}

\begin{figure}[h]
  \centering
  \includegraphics{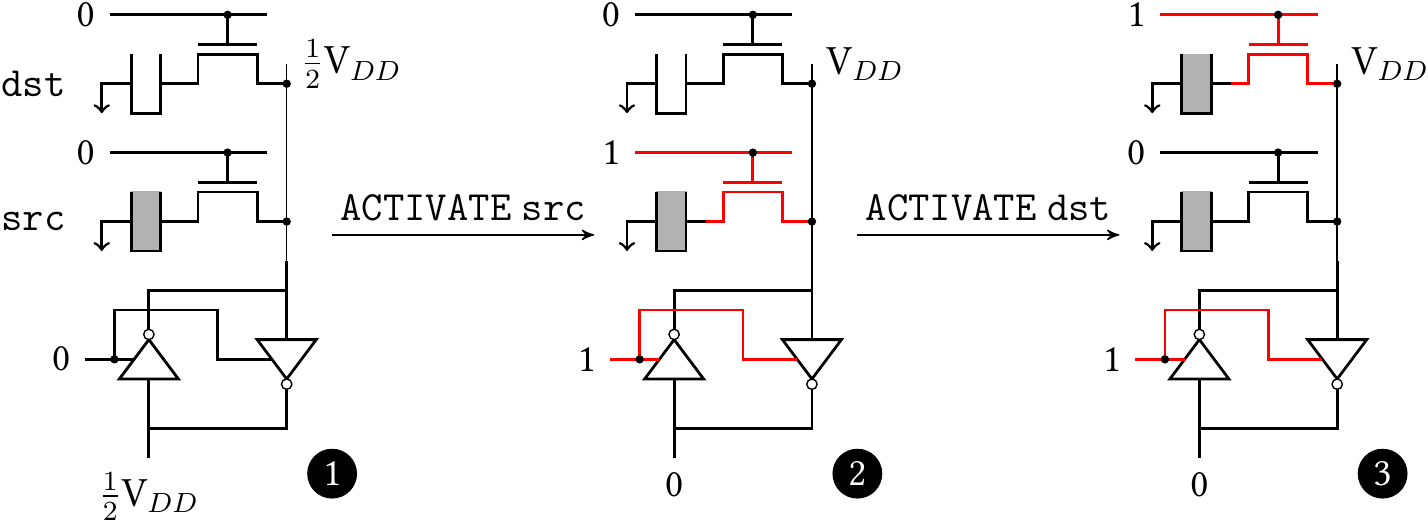}
  \caption{RowClone: Fast Parallel Mode}
  \label{fig:cell-fpm}
\end{figure}

While the first two observations are direct implications from the
design of commodity DRAM, the third observation exploits the fact
that DRAM cells can cause only a small perturbation on the bitline
voltage. Figure~\ref{fig:cell-fpm} pictorially shows how this
observation can be used to copy data between two cells that share
a sense amplifier.

The figure shows two cells (\src and \dst) connected to a single
sense amplifier. In the initial state, we assume that \src is
fully charged and \dst is fully empty, and the sense amplifier is
in the precharged state (\ding{202}). In this state, FPM issues an
\cmdact to \src. At the end of the activation operation, the sense
amplifier moves to a stable state where the bitline is at a
voltage level of \vdd and the charge in \src is fully restored
(\ding{203}). FPM follows this operation with an \cmdact to \dst,
without an intervening \cmdpre. This operation lowers the wordline
corresponding to \src and raises the wordline of \dst, connecting
\dst to the bitline.  Since the bitline is already fully
activated, even though \dst is initially empty, the perturbation
caused by the cell is not sufficient to flip the state of the
bitline. As a result, the sense amplifier continues to drive the
bitline to \vdd, thereby pushing \dst to a fully charged state
(\ding{204}).

It can be shown that regardless of the initial state of \src and
\dst, the above operation copies the data from \src to
\dst. Given that each \cmdact operates on an entire row of DRAM
cells, the above operation can copy multiple kilo bytes of data
with just two back-to-back \cmdact operations.

Unfortunately, modern DRAM chips do not allow another \cmdact to
an already activated bank -- the expected result of such an action
is undefined. This is because a modern DRAM chip allows at most
one row (subarray) within each bank to be activated.  If a bank
that already has a row (subarray) activated receives an \cmdact to
a different subarray, the currently activated subarray must first
be precharged~\cite{salp}. Some DRAM manufacturers design their
chips to drop back-to-back {\cmdact}s to the same bank.

To support FPM, RowClone changes the way a DRAM chip handles
back-to-back {\cmdact}s to the same bank. When an already
activated bank receives an \cmdact to a row, the chip allows the
command to proceed if and only if the command is to a row that
belongs to the currently activated subarray. If the row does not
belong to the currently activated subarray, then the chip takes
the action it normally does with back-to-back {\cmdact}s---e.g.,
drop it.  Since the logic to determine the subarray corresponding
to a row address is already present in today's chips, implementing
FPM only requires a comparison to check if the row address of an
\cmdact belongs to the currently activated subarray, the cost of
which is almost negligible.

\textbf{Summary.} To copy data from \src to \dst within the same
subarray, FPM first issues an \cmdact to \src. This copies the
data from \src to the subarray row buffer. FPM then issues an
\cmdact to \dst. This modifies the input to the subarray
row-decoder from \src to \dst and connects the cells of \dst row
to the row buffer. This, in effect, copies the data from the sense
amplifiers to the destination row.  With these two steps, FPM can
copy a 4KB page of data 12.0x faster and with 74.4x less energy
than an existing system (we describe the methodology in
Section~\ref{sec:lte}).

\textbf{Limitations.} FPM has two constraints that limit its
general applicability. First, it requires the source and
destination rows to be within the same subarray (i.e., share the
same set of sense amplifiers). Second, it cannot partially copy
data from one row to another. Despite these limitations, FPM can
be immediately applied to today's systems to accelerate two
commonly used primitives in modern systems -- Copy-on-Write and
Bulk Zeroing.  In the following section, we describe the second
mode of RowClone -- the Pipelined Serial Mode (PSM). Although not
as fast or energy-efficient as FPM, PSM addresses these two
limitations of FPM.

\subsection{Pipelined Serial Mode}
\label{sec:rowclone-psm}

The Pipelined Serial Mode efficiently copies data from a source
row in one bank to a destination row in a \emph{different}
bank. PSM exploits the fact that a single internal bus that is
shared across all the banks is used for both read and write
operations. This enables the opportunity to copy an arbitrary
quantity of data one cache line at a time from one bank to another
in a pipelined manner.

To copy data from a source row in one bank to a destination row in
a different bank, PSM first activates the corresponding rows in
both banks. It then puts the source bank into \emph{read mode},
the destination bank into \emph{write mode}, and transfers data
one cache line (corresponding to a column of data---64 bytes) at a
time. For this purpose, RowClone introduces a new DRAM command
called \cmdtr. The \cmdtr command takes four parameters: 1)~source
bank index, 2)~source column index, 3)~destination bank index, and
4)~destination column index. It copies the cache line
corresponding to the source column index in the activated row of
the source bank to the cache line corresponding to the destination
column index in the activated row of the destination bank.

\begin{figure}[h]
  \centering
  \includegraphics[angle=90]{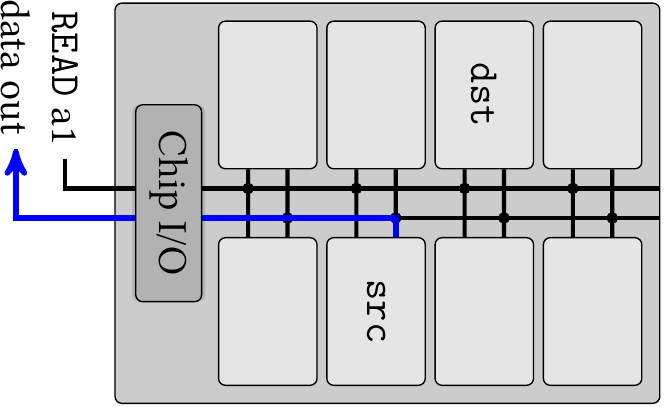}
  \includegraphics[angle=90]{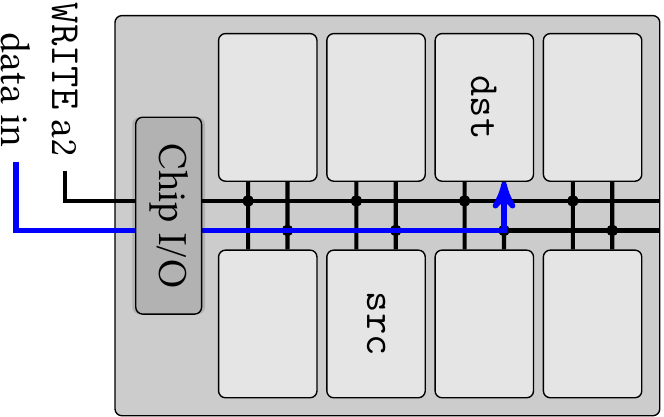}
  \includegraphics[angle=90]{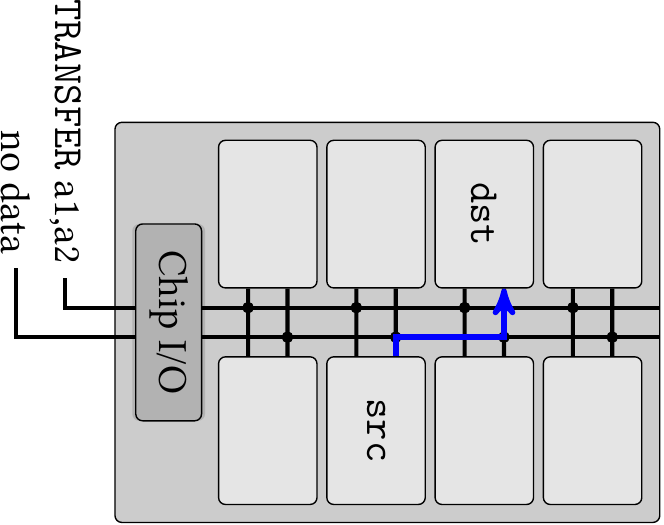}
  \caption{RowClone: Pipelined Serial Mode}
  \label{fig:psm}
\end{figure}

Unlike \cmdrd/\cmdwr, which interact with the memory channel
connecting the processor and main memory, \cmdtr does not transfer
data outside the chip. Figure~\ref{fig:psm} pictorially compares
the operation of the \cmdtr command with that of \cmdrd and
\cmdwr. The dashed lines indicate the data flow corresponding to
the three commands. As shown in the figure, in contrast to the
\cmdrd or \cmdwr commands, \cmdtr does not transfer data from or
to the memory channel.

\subsection{Mechanism for Bulk Data Copy}
\label{sec:bulk-copy}

When the data from a source row (\src) needs to be copied to a
destination row (\dst), there are three possible cases depending
on the location of \src and \dst: 1)~\src and \dst are within the
same subarray, 2)~\src and \dst are in different banks, 3)~\src
and \dst are in different subarrays within the same bank. For case
1 and case 2, RowClone uses FPM and PSM, respectively, to complete
the operation (as described in Sections~\ref{sec:rowclone-fpm} and
\ref{sec:rowclone-psm}).

For the third case, when \src and \dst are in different subarrays
within the same bank, one can imagine a mechanism that uses the
global bitlines (shared across all subarrays within a bank --
described in \cite{salp}) to copy data across the two rows in
different subarrays. However, RowClone does not employ such a
mechanism for two reasons.  First, it is not possible in today's
DRAM chips to activate multiple subarrays within the same bank
simultaneously.  Second, even if we enable simultaneous activation
of multiple subarrays, as in~\cite{salp}, transferring data from
one row buffer to another using the global bitlines requires the
bank I/O circuitry to switch between read and write modes for each
cache line transfer. This switching incurs significant latency
overhead.  To keep the design simple, for such an intra-bank copy
operation, RowClone uses PSM to first copy the data from \src
to a temporary row (\tmp) in a different bank. It then uses PSM
again to copy the data from \tmp to \dst. The capacity lost
due to reserving one row within each bank is negligible (0.0015\%
for a bank with 64k rows).

Despite its location constraints, FPM can be used to accelerate
\emph{Copy-on-Write} (CoW), an important primitive in modern
systems.  CoW is used by most modern operating systems (OS) to
postpone an expensive copy operation until it is actually
needed. When data of one virtual page needs to be copied to
another, instead of creating a copy, the OS points both virtual
pages to the same physical page (source) and marks the page as
read-only. In the future, when one of the sharers attempts to
write to the page, the OS allocates a new physical page
(destination) for the writer and copies the contents of the source
page to the newly allocated page. Fortunately, prior to allocating
the destination page, the OS already knows the location of the
source physical page. Therefore, it can ensure that the
destination is allocated in the same subarray as the source,
thereby enabling the processor to use FPM to perform the copy.

\subsection{Mechanism for Bulk Data Initialization}
\label{sec:bulk-initialization}

Bulk data initialization sets a large block of memory to a
specific value. To perform this operation efficiently, RowClone
first initializes a single DRAM row with the corresponding
value. It then uses the appropriate copy mechanism (from
Section~\ref{sec:bulk-copy}) to copy the data to the other rows to
be initialized.

Bulk Zeroing (or BuZ), a special case of bulk initialization, is a
frequently occurring operation in today's
systems~\cite{bulk-copy-initialize,why-nothing-matters}. To
accelerate BuZ, one can reserve one row in each subarray that is
always initialized to zero. By doing so, RowClone can use FPM to
efficiently BuZ any row in DRAM by copying data from the reserved
zero row of the corresponding subarray into the destination
row. The capacity loss of reserving one row out of 512 rows in
each subarray is very modest (0.2\%).

While the reserved rows can potentially lead to gaps in the
physical address space, we can use an appropriate memory
interleaving technique that maps consecutive rows to different
subarrays. Such a technique ensures that the reserved zero rows
are contiguously located in the physical address space. Note that
interleaving techniques commonly used in today's systems (e.g.,
row or cache line interleaving) have this property.

\section{In-DRAM Bulk AND and OR}
\label{sec:and-or}

In this section, we describe In-DRAM AND/OR (IDAO), which is a
mechanism to perform bulk bitwise AND and OR operations completely
inside DRAM. In addition to simple masking and initialization
tasks, these operations are useful in important data structures
like bitmap indices. For example, bitmap
indices~\cite{bmide,bmidc} can be more efficient than
commonly-used B-trees for performing range queries and joins in
databases~\cite{bmide,fastbit,bicompression}. In fact, bitmap
indices are supported by many real-world database implementations
(e.g., Redis~\cite{redis}, Fastbit~\cite{fastbit}). Improving the
throughput of bitwise AND and OR operations can boost the
performance of such bitmap indices.

\subsection{Mechanism}
\label{sec:bitwise-and-or}

As described in Section~\ref{sec:cell-operation}, when a DRAM cell
is connected to a bitline precharged to \halfvdd, the cell induces
a deviation on the bitline, and the deviation is amplified by the
sense amplifier. IDAO exploits the following fact about DRAM
cell operation.
\begin{quote}
  The final state of the bitline after amplification is determined
  solely by the deviation on the bitline after the charge sharing
  phase (after state \ding{204} in
  Figure~\ref{fig:cell-operation}). If the deviation is positive
  (i.e., towards \vdd), the bitline is amplified to
  \vdd. Otherwise, if the deviation is negative (i.e., towards
  $0$), the bitline is amplified to $0$.
\end{quote}

\subsubsection{Triple-Row Activation}
\label{sec:triple-row-activation}

IDAO simultaneously connects three cells as opposed to a single
cell to a sense amplifier. When three cells are connected to the
bitline, the deviation of the bitline after charge sharing is
determined by the \emph{majority value} of the three
cells. Specifically, if at least two cells are initially in the
charged state, the effective voltage level of the three cells is
at least \ttvdd. This results in a positive deviation on the
bitline. On the other hand, if at most one cell is initially in
the charged state, the effective voltage level of the three cells
is at most \otvdd. This results in a negative deviation on the
bitline voltage. As a result, the final state of the bitline is
determined by the logical majority value of the three cells.

\begin{figure}[h]
  \centering
  \includegraphics[scale=1.4]{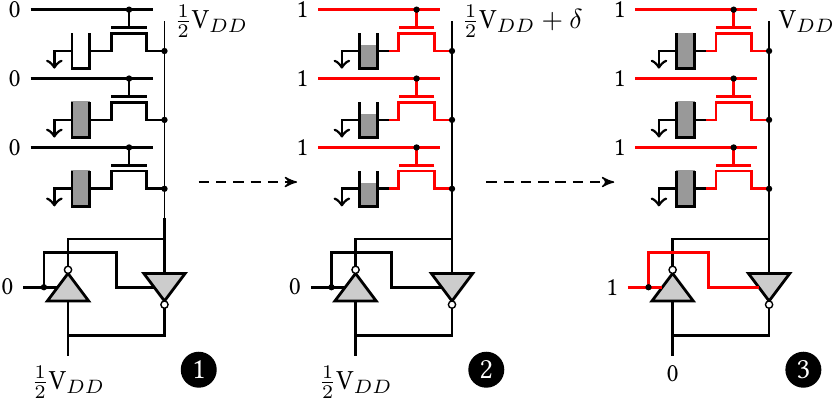}
  \caption[Triple-row activation in DRAM]{Triple-row activation}
  \label{fig:triple-row-activation}
\end{figure}

Figure~\ref{fig:triple-row-activation} shows an example of
activating three cells simultaneously. In the figure, we assume
that two of the three cells are initially in the charged state and
the third cell is in the empty state \ding{202}. When the
wordlines of all the three cells are raised simultaneously
\ding{203}, charge sharing results in a positive deviation on the
bitline. Hence, after sense amplification, the sense amplifier
drives the bitline to \vdd and fully charges all three cells
\ding{204}.

More generally, if the cell's capacitance is $C_c$, the the
bitline's is $C_b$, and if $k$ of the three cells are initially in
the charged state, based on the charge sharing
principles~\cite{dram-cd}, the deviation $\delta$ on the bitline
voltage level is given by,
\begin{eqnarray}
  \delta &=& \frac{k.C_c.V_{DD} + C_b.\frac{1}{2}V_{DD}}{3C_c + C_b}
  - \frac{1}{2}V_{DD}\nonumber\\
  &=& \frac{(2k - 3)C_c}{6C_c + 2C_b}V_{DD}\label{eqn:delta}
\end{eqnarray}
From the above equation, it is clear that $\delta$ is positive for
$k = 2,3$, and $\delta$ is negative for $k = 0,1$. Therefore,
after amplification, the final voltage level on the bitline is
\vdd for $k = 2,3$ and $0$ for $k = 0,1$.

If $A$, $B$, and $C$ represent the logical values of the three
cells, then the final state of the bitline is $AB + BC + CA$ (i.e., at
least two of the values should be $1$ for the final state to be
$1$). Importantly, using simple boolean algebra, this expression
can be rewritten as $C(A + B) + \overline{C}(AB)$. In other words,
if the initial state of $C$ is $1$, then the final state of the
bitline is a bitwise OR of $A$ and $B$. Otherwise, if the initial
state of $C$ is $0$, then the final state of the bitline is a
bitwise AND of $A$ and $B$. Therefore, by controlling the value of
the cell C, we can execute a bitwise AND or bitwise OR operation
of the remaining two cells using the sense amplifier. Due to the
regular bulk operation of cells in DRAM, this approach naturally
extends to an entire row of DRAM cells and sense amplifiers,
enabling a multi-kilobyte-wide bitwise AND/OR
operation.

\subsubsection{Challenges}
\label{sec:and-or-challenges}

There are two challenges in this approach. First,
Equation~\ref{eqn:delta} assumes that the cells involved in the
triple-row activation are either fully charged or fully
empty. However, DRAM cells leak charge over time. Therefore, the
triple-row activation may not operate as expected. This problem
may be exacerbated by process variation in DRAM cells. Second, as
shown in Figure~\ref{fig:triple-row-activation}
(state~\ding{204}), at the end of the triple-row activation, the
data in all the three cells are overwritten with the final state
of the bitline. In other words, this approach overwrites the
source data with the final value. In the following sections, we
describe a simple implementation of IDAO that addresses these
challenges.

\subsubsection{Implementation of IDAO}
\label{sec:and-or-mechanism}

To ensure that the source data does not get modified, IDAO first
\emph{copies} the data from the two source rows to two reserved
temporary rows ($T1$ and $T2$). Depending on the operation to be
performed (AND or OR), our mechanism initializes a third reserved
temporary row $T3$ to ($0$ or $1$). It then simultaneously
activates the three rows $T1$, $T2$, and $T3$. It finally copies
the result to the destination row. For example, to perform a
bitwise AND of two rows $A$ and $B$ and store the result in row
$R$, IDAO performs the following steps.
\begin{enumerate}\itemsep0pt\parsep0pt\parskip0pt
\item \emph{Copy} data of row $A$ to row $T1$
\item \emph{Copy} data of row $B$ to row $T2$
\item \emph{Initialize} row $T3$ to $0$
\item \emph{Activate} rows $T1$, $T2$, and $T3$ simultaneously
\item \emph{Copy} data of row $T1$ to row $R$
\end{enumerate}

While the above mechanism is simple, the copy operations, if
performed naively, will nullify the benefits of our
mechanism. Fortunately, IDAO uses RowClone (described in
Section~\ref{sec:rowclone}), to perform row-to-row copy operations
quickly and efficiently within DRAM. To recap, RowClone-FPM copies
data within a subarray by issuing two back-to-back {\cmdact}s to
the source row and the destination row, without an intervening
\cmdpre. RowClone-PSM efficiently copies data between two banks by
using the shared internal bus to overlap the read to the source
bank with the write to the destination bank.

With RowClone, all three copy operations (Steps 1, 2, and 5) and
the initialization operation (Step 3) can be performed efficiently
within DRAM. To use RowClone for the initialization operation,
IDAO reserves two additional rows, $C0$ and $C1$. $C0$ is
pre-initialized to $0$ and $C1$ is pre-initialized to 1. Depending
on the operation to be performed, our mechanism uses RowClone to
copy either $C0$ or $C1$ to $T3$. Furthermore, to maximize the use
of RowClone-FPM, IDAO reserves five rows in each subarray to serve
as the temporary rows ($T1$, $T2$, and $T3$) and the control rows
($C0$ and $C1$).

In the best case, when all the three rows involved in the
operation ($A$, $B$, and $R$) are in the same subarray, IDAO can
use RowClone-FPM for all copy and initialization
operations. However, if the three rows are in different
banks/subarrays, some of the three copy operations have to use
RowClone-PSM. In the worst case, when all three copy operations
have to use RowClone-PSM, IDAO will consume higher latency than
the baseline. However, when only one or two RowClone-PSM
operations are required, IDAO will be faster and more
energy-efficient than existing systems. As our goal in this
article is to demonstrate the power of our approach, in the rest
of the article, we will focus our attention on the case when all
rows involved in the bitwise operation are in the same subarray.

\subsubsection{Reliability of Our Mechanism}                                                                                                                             
While the above implementation trivially addresses the second
challenge (modification of the source data), it also addresses the
first challenge (DRAM cell leakage). This is because, in our
approach, the source (and the control) data are copied to the rows
$T1$, $T2$ and $T3$ \emph{just} before the triple-row
activation. Each copy operation takes much less than $1~{\mu}s$,
which is five \emph{orders} of magnitude less than the typical
refresh interval ($64~ms$). Consequently, the cells involved in
the triple-row activation are very close to the fully refreshed
state before the operation, thereby ensuring reliable operation of
the triple-row activation. Having said that, an important aspect
of the implementation is that a chip that fails the tests for
triple-row activation (e.g., due to process variation) \emph{can
  still be used as a regular DRAM chip}. As a result, this
approach is likely to have little impact on the overall yield of
DRAM chips, which is a major concern for manufacturers.

\subsubsection{Latency Optimization}

To complete an intra-subarray copy, RowClone-FPM uses two
{\cmdact}s (back-to-back) followed by a \cmdpre
operation. Assuming typical DRAM timing parameters (tRAS = $35ns$
and tRP = $15ns$), each copy operation consumes $85ns$. As IDAO is
essentially four RowClone-FPM operations (as described in the
previous section), the overall latency of a bitwise AND/OR
operation is $4 \times 85ns = 340ns$.

In a RowClone-FPM operation, although the second \cmdact does not
involve any sense amplification (the sense amplifiers are already
activated), the RowClone paper~\cite{rowclone} assumes the \cmdact
consumes the full tRAS latency. However, by controlling the rows
$D1$, $D2$, and $D3$ using a separate row decoder, it is possible
to overlap the \cmdact to the destination fully with the \cmdact
to the source row, by raising the wordline of the destination row
towards the end of the sense amplification of the source row. This
mechanism is similar to the inter-segment copy operation described
in Tiered-Latency DRAM~\cite{tl-dram} (Section 4.4). With this
aggressive mechanism, the latency of a RowClone-FPM operation
reduces to $50ns$ (one \cmdact and one \cmdpre). Therefore, the
overall latency of a bitwise AND/OR operation is $200ns$. We will
refer to this enhanced mechanism as \emph{aggressive}, and the
approach that uses the simple back-to-back \cmdact operations as
\emph{conservative}.

\section{End-to-end System Support}
\label{sec:system-support}

Both RowClone and IDAO are substrates that exploit DRAM technology
to perform bulk copy, initialization, and bitwise AND/OR
operations efficiently inside DRAM. However, to exploit these
substrates, we need support from the rest of the layers in the
system stack, namely, the instruction set architecture, the
microarchitecture, and the system software. In this section, we
describe this support in detail.

\subsection{ISA Support}
\label{sec:isa-changes}

To enable the software to communicate occurrences of the bulk
operations to the hardware, the mechanisms introduce four new
instructions to the ISA: \mcpy, \minit, \band, and \bor.
Table~\ref{tab:isa-semantics} describes the semantics of these four
new instructions.  The mechanisms deliberately keep the
semantics of the instructions simple in order to relieve the
software from worrying about microarchitectural aspects of the
DRAM substrate such as row size, alignment, etc.~(discussed in
Section~\ref{sec:offset-alignment-size}). Note that such
instructions are already present in some of the instructions sets
in modern processors -- e.g., \texttt{rep movsd}, \texttt{rep
  stosb}, \texttt{ermsb} in x86~\cite{x86-ermsb} and \texttt{mvcl}
in IBM S/390~\cite{s390}.

\begin{table}[h]
  \centering
  \begin{tabular}{llp{3in}}
  \toprule
  Instruction & Operands & Semantics\\
  \midrule
  \mcpy & \emph{src, dst, size} & Copy \emph{size} bytes from
  \emph{src} to \emph{dst}\\
  \midrule
  \minit & \emph{dst, size, val} & Set \emph{size} bytes to
  \emph{val} at \emph{dst}\\
  \midrule
  \multirow{3}{*}{\band} & \multirow{3}{*}{\emph{src1, src2, dst, size}} & Perform bitwise AND of
  \emph{size} bytes of \emph{src1} with \emph{size} bytes of \emph{src2} and store the result
  in the \emph{dst}\\
  \midrule
  \multirow{3}{*}{\bor} & \multirow{3}{*}{\emph{src1, src2, dst, size}} & Perform bitwise OR of
  \emph{size} bytes of \emph{src1} with \emph{size} bytes of \emph{src2} and store the result
  in the \emph{dst}\\
  \bottomrule
\end{tabular}

  \caption{Semantics of the \mcpy, \minit, \band, and \bor
    instructions}
  \label{tab:isa-semantics}
\end{table}

There are three points to note regarding the execution semantics
of these operations. First, the processor does not guarantee
atomicity for any of these instructions, but note that existing
systems also do not guarantee atomicity for such
operations. Therefore, the software must take care of atomicity
requirements using explicit synchronization. However, the
microarchitectural implementation ensures that any data in the
on-chip caches is kept consistent during the execution of these
operations (Section~\ref{sec:rowclone-cache-coherence}). Second,
the processor handles any page faults during the execution of
these operations. Third, the processor can take interrupts during
the execution of these operations.

\subsection{Processor Microarchitecture Support}
\label{sec:uarch-changes}

The microarchitectural implementation of the new instructions has
two parts. The first part determines if a particular instance of
the instructions can be fully/partially accelerated by
RowClone/IDAO. The second part involves the changes required to
the cache coherence protocol to ensure coherence of data in the
on-chip caches. We discuss these parts in this section.

\subsubsection{Source/Destination Alignment and Size}
\label{sec:offset-alignment-size}

For the processor to accelerate a copy/initialization operation
using RowClone, the operation must satisfy certain alignment and
size constraints. Specifically, for an operation to be accelerated
by FPM, 1)~the source and destination regions should be within the
same subarray, 2)~the source and destination regions should be
row-aligned, and 3)~the operation should span an entire row. On
the other hand, for an operation to be accelerated by PSM, the
source and destination regions should be cache line-aligned and the
operation must span a full cache line.

Upon encountering a \mcpy/\minit instruction, the processor
divides the region to be copied/initialized into three portions:
1)~row-aligned row-sized portions that can be accelerated using
FPM, 2)~cache line-aligned cache line-sized portions that can be
accelerated using PSM, and 3)~the remaining portions that can be
performed by the processor. For the first two regions, the
processor sends appropriate requests to the memory controller,
which completes the operations and sends an acknowledgment back to
the processor. Since \cmdtr copies only a single cache line, a
bulk copy using PSM can be interleaved with other commands to
memory. The processor completes the operation for the third region
similarly to how it is done in today's systems. Note that the CPU
can offload all these operations to the memory controller. In such
a design, the CPU need not be made aware of the DRAM organization
(e.g., row size and alignment, subarray mapping, etc.).

For each instance of \band/\bor instruction, the processor follows
a similar procedure. However, only the row-aligned row-sized
portions are accelerated using IDAO. The remaining portions are
still performed by the CPU. For the row-aligned row-sized regions,
some of the copy operations may require RowClone-PSM. For each row
of the operation, the processor determines if the number of
RowClone-PSM operations required is three. If so, the processor
completes the execution in the CPU. Otherwise, the operation is
completed using IDAO.

\subsubsection{Managing On-Chip Cache Coherence}
\label{sec:rowclone-cache-coherence}

Both RowClone and IDAO allow the memory controller to directly
read/modify data in memory without going through the on-chip
caches. Therefore, to ensure cache coherence, the controller
appropriately handles cache lines from the source and destination
regions that may be present in the caches before issuing the
in-DRAM operations to memory.

First, the memory controller writes back any dirty cache line from
the source region as the main memory version of such a cache line
is likely stale. Using the data in memory before flushing such
cache lines will lead to stale data being copied to the
destination region. Second, the controller invalidates any cache
line (clean or dirty) from the destination region that is cached
in the on-chip caches. This is because after performing the
operation, the cached version of these blocks may contain stale
data. The controller already has the ability to perform such
flushes and invalidations to support Direct Memory Access
(DMA)~\cite{intel-dma}. After performing the necessary flushes and
invalidations, the memory controller performs the in-DRAM
operation. To ensure that cache lines of the destination region
are not cached again by the processor in the meantime, the memory
controller blocks all requests (including prefetches) to the
destination region until the copy or initialization operation is
complete. A recent work, LazyPIM~\cite{lazypim}, proposes an
approach to perform the coherence operations lazily by comparing
the signatures of data that were accessed in memory and the data
that is cached on-chip. Our mechanisms can be combined with such
works.

For RowClone, while performing the flushes and invalidates as
mentioned above will ensure coherence, we propose a modified
solution to handle dirty cache lines of the source region to
reduce memory bandwidth consumption. When the memory controller
identifies a dirty cache line belonging to the source region while
performing a copy, it creates an in-cache copy of the source cache
line with the tag corresponding to the destination cache
line. This has two benefits. First, it avoids the additional
memory flush required for the dirty source cache line. Second and
more importantly, the controller does not have to wait for all the
dirty source cache lines to be flushed before it can perform the
copy. In the evaluation section, we will describe another
optimization, called RowClone-Zero-Insert, which inserts clean
zero cache lines into the cache to further optimize Bulk Zeroing.
This optimization does not require further changes to the proposed
modifications to the cache coherence protocol.

Although the two mechanisms require the controller to manage cache
coherence, it does not affect memory consistency --- i.e., the
ordering of accesses by concurrent readers and/or writers to the
source or destination regions. As mentioned before, such an
operation is not guaranteed to be atomic even in current systems,
and software needs to perform the operation within a critical
section to ensure atomicity.

\subsection{Software Support}
\label{sec:os-changes}

The minimum support required from the system software is the use
of the proposed instructions to indicate bulk data operations to
the processor. Although one can have a working system with just
this support, the maximum latency and energy benefits can be
obtained if the hardware is able to accelerate most operations
using FPM rather than PSM. Increasing the likelihood of the use of
the FPM mode requires further support from the operating system
(OS) on two aspects: 1)~page mapping, and 2)~granularity of the
operation.

\subsubsection{Subarray-Aware Page Mapping}
\label{sec:subarray-awareness}

The use of FPM requires the source row and the destination row of
a copy operation to be within the same subarray. Therefore, to
maximize the use of FPM, the OS page mapping algorithm should be
aware of subarrays so that it can allocate a destination page of a
copy operation in the same subarray as the source page. More
specifically, the OS should have knowledge of which pages map to
the same subarray in DRAM. We propose that DRAM expose this
information to software using the small EEPROM that already exists
in today's DRAM modules. This EEPROM, called the Serial Presence
Detect (SPD)~\cite{spd}, stores information about the DRAM chips
that is read by the memory controller at system bootup. Exposing
the subarray mapping information will require only a few
additional bytes to communicate the bits of the physical address
that map to the subarray index. To increase DRAM yield, DRAM
manufacturers design chips with spare rows that can be mapped to
faulty rows~\cite{spare-row-mapping}. The mechanisms can work with
this technique by either requiring that each faulty row is
remapped to a spare row within the same subarray, or exposing the
location of all faulty rows to the memory controller so that it
can use PSM to copy data across such rows.

Once the OS has the mapping information between physical pages and
subarrays, it maintains multiple pools of free pages, one pool for
each subarray. When the OS allocates the destination page for a
copy operation (e.g., for a \emph{Copy-on-Write} operation), it
chooses the page from the same pool (subarray) as the source
page. Note that this approach does not require contiguous pages to
be placed within the same subarray. As mentioned before, commonly
used memory interleaving techniques spread out contiguous pages
across as many banks/subarrays as possible to improve
parallelism. Therefore, both the source and destination of a bulk
copy operation can be spread out across many subarrays.

\pagebreak
\subsubsection{Granularity of the Operations}
\label{sec:memory-interleaving}

The second aspect that affects the use of FPM and IDAO is the
granularity at which data is copied or initialized. These
mechanisms have a minimum granularity at which they operate. There
are two factors that affect this minimum granularity: 1)~the size
of each DRAM row, and 2)~the memory interleaving employed by the
controller.

First, in each chip, these mechanisms operate on an entire row of
data. Second, to extract maximum bandwidth, some memory
interleaving techniques map consecutive cache lines to different
memory channels in the system. Therefore, to operate on a
contiguous region of data with such interleaving strategies, the
mechanisms must perform the operation in each channel. The minimum
amount of data in such a scenario is the product of the row size
and the number of channels.

To maximize the likelihood of using FPM and IDAO, the system or
application software must ensure that the region of data involved
in the operation is at least as large as this minimum
granularity. For this purpose, we propose to expose this minimum
granularity to the software through a special register, which we
call the \emph{Minimum DRAM Granularity Register} (MDGR). On
system bootup, the memory controller initializes the MCGR based on
the row size and the memory interleaving strategy, which can later
be used by the OS for effectively exploiting RowClone/IDAO. Note
that some previously proposed techniques such as sub-wordline
activation~\cite{rethinking-dram} or
mini-rank~\cite{threaded-module,mini-rank} can be combined with
our mechanisms to reduce the minimum granularity.

\section{Evaluation}
\label{sec:evaluation}

To highlight the benefits of performing various operations
completely inside DRAM, we first compare the raw latency and
energy required to perform these operations using different
mechanisms (Section~\ref{sec:lte}). We then present the
quantitative evaluation of applications for RowClone and IDAO in
Sections~\ref{sec:rowclone-eval} and \ref{sec:idao-eval},
respectively.

\subsection{Latency and Energy Analysis}
\label{sec:lte}

We estimate latency using DDR3-1600 timing parameters. We estimate
energy using the Rambus power model~\cite{rambus-power}. Our
energy calculations only include the energy consumed by the DRAM
module and the DRAM channel Table~\ref{tab:latency-energy} shows
the latency and energy consumption due to the different mechanisms
for bulk copy, zero, and bitwise AND/OR operations. The table also
shows the potential reduction in latency and energy by performing
these operations completely inside DRAM.

\begin{table}[h]
  \centering
  \begin{tabular}{crrrrr}
  \toprule
  & \multirow{4}{*}{\textbf{Mechanism}} &
  \multicolumn{2}{c}{\textbf{Absolute}} &
  \multicolumn{2}{c}{\textbf{Reduction}}\\
  & & & \small{Memory} & & \small{Memory}\vspace{-1mm}\\ 
  & & \small{Latency}  & \small{Energy} & \small{Latency} & \small{Energy}\\
  & & (ns) & ($\mu$J)\\
  \toprule
  \multirow{5}{*}{\textbf{Copy}} & {Baseline} & 1020 & 3.6 & 1.00x & 1.0x\\
  \cmidrule{2-6}
  & {FPM} & 85 & 0.04 & \textbf{12.0x} & \textbf{74.4x}\\
  \cmidrule{2-6}
  & {Inter-Bank - PSM} & 510 & 1.1 & 2.0x & 3.2x\\
  \cmidrule{2-6}
  & {Intra-Bank - PSM} & 1020 & 2.5 & 1.0x & 1.5x\\
  \toprule
  \multirow{2}{*}{\textbf{Zero}} & {Baseline} & 510 & 2.0 & 1.00x  & 1.0x\\
  \cmidrule{2-6}
  & {FPM} & 85 & 0.05 & \textbf{6.0x} & \textbf{41.5x}\\
  \toprule
  \multirow{4}{*}{\textbf{AND/OR}} & {Baseline} & 1530 & 5.0 & 1.00x
  & 1.0x\\ 
  \cmidrule{2-6}
  & {IDAO-Conservative} & 320 & 0.16 & 4.78x & 31.6x \\
  \cmidrule{2-6}
  & {IDAO-Aggressive} & 200 & 0.10 & \textbf{7.65x} & \textbf{50.5x} \\
  \bottomrule
\end{tabular}

  \caption{DRAM latency and memory energy reductions}
  \label{tab:latency-energy}
\end{table}

First, for bulk copy operations, RowClone-FPM reduces latency by
12x and energy consumption by 74.4x compared to existing
interfaces. While PSM does not provide as much reduction as FPM,
PSM still reduces latency by 2x and energy consumption by 3.2x
compared to the baseline. Second, for bulk zeroing operations,
RowClone can always use the FPM mode as it reserves a single zero
row in each subarray of DRAM. As a result, it can reduce the
latency of bulk zeroing by 6x and energy consumption by 41.5x
compared to existing interfaces. Finally, for bitwise AND/OR
operations, even with conservative timing parameters, IDAO can
reduce latency by 4.78x and energy consumption by 31.6x. With more
aggressive timing parameters, IDAO reduces latency by 7.65x and
energy by 50.5x.

The improvement in sustained throughput due to RowClone and IDAO
for the respective operations is similar to the improvements in
latency. The main takeaway from these results is that, for systems
that use DRAM to store majority of their data (which includes most
of today's systems), these in-DRAM mechanisms are probably the
best performing and the most energy-efficient way of performing
the respective operations. We will now provide quantitative
evaluation of these mechanisms on some real applications.

\subsection{Applications for RowClone}
\label{sec:rowclone-eval}

Our evaluations use an in-house cycle-level multi-core simulator
similar to memsim~\cite{eaf,icp,memsim} along with a
cycle-accurate command-level DDR3 DRAM simulator, similar to
Ramulator~\cite{ramulator,ramulator-github}. The multi-core
simulator models out-of-order cores, each with a private
last-level cache. We evaluate the benefits of RowClone using 1)~a
case study of the \fork system call, an important operation used
by modern operating systems, 2)~six copy and initialization
intensive benchmarks: \emph{bootup}, \emph{compile},
\emph{forkbench}, \emph{memcached}~\cite{memcached},
\emph{mysql}~\cite{mysql}, and \emph{shell}
(Section~\ref{res:rowclone-ii-apps} describes these benchmarks),
and 3)~a wide variety of multi-core workloads comprising the
copy/initialization intensive applications running alongside
memory-intensive applications from the SPEC CPU2006 benchmark
suite~\cite{spec2006}. Note that benchmarks such as SPEC CPU2006,
which predominantly stress the CPU, typically use a small number
of page copy and initialization operations and therefore would
serve as poor individual evaluation benchmarks for RowClone.

We collected instruction traces for our workloads using
Bochs~\cite{bochs}, a full-system x86-64 emulator, running a
GNU/Linux system. We modify the kernel's implementation of page
copy/initialization to use the \mcpy and \minit instructions and
mark these instructions in our traces. For the \fork benchmark, we
used the Wind River Simics full system simulator~\cite{simics} to
collect the traces. We collect 1-billion instruction traces of the
representative portions of these workloads. We use the instruction
throughput (IPC) metric to measure single-core performance. We
evaluate multi-core runs using the weighted speedup
metric~\cite{weighted-speedup,weighted-speedup-2}. This metric is
used by many prior works
(e.g.,~\cite{dbi,hat,bliss,bliss-tpds,rbla,row-buffer-nvm,mlc-pcm,multi-pref-mc})
to measure system throughput for multi-programmed workloads. In
addition to weighted speedup, we use five other
performance/fairness/bandwidth/energy metrics, as shown in
Table~\ref{tab:multi-core-ws}.

\subsubsection{The {\large{\fork}} System Call}
\label{sec:res-fork}

\fork is one of the most expensive yet frequently-used system
calls in modern systems~\cite{fork-exp}. Since \fork triggers a
large number of CoW operations (as a result of updates to shared
pages from the parent or child process), RowClone can
significantly improve the performance of \fork.

The performance of \fork depends on two parameters: 1)~the size of
the address space used by the parent---which determines how much
data may potentially have to be copied, and 2)~the number of pages
updated after the \fork operation by either the parent or the
child---which determines how much data is actually copied. To
exercise these two parameters, we create a microbenchmark,
\forkbench, which first creates an array of size \ubsize and
initializes the array with random values. It then forks
itself. The child process updates $N$ random pages (by updating a
cache line within each page) and exits; the parent process waits
for the child process to complete before exiting itself.

As such, we expect the number of copy operations to depend on
$N$---the number of pages copied. Therefore, one may expect
RowClone's performance benefits to be proportional to $N$.
However, an application's performance typically depends on the
{\em overall memory access rate}~\cite{mise,asm}, and RowClone can
only improve performance by reducing the {\em memory access rate
  due to copy operations}. As a result, we expect the performance
improvement due to RowClone to primarily depend on the
\emph{fraction} of memory traffic (total bytes transferred over
the memory channel) generated by copy operations. We refer to this
fraction as FMTC---Fraction of Memory Traffic due to Copies.

\begin{figure}[h]
  \centering
  \includegraphics{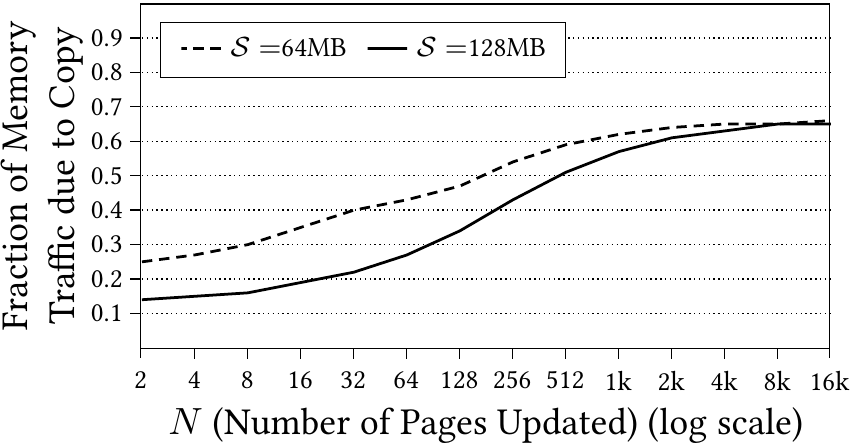}
  \caption[Memory traffic due to copy in \forkbench]{FMTC of
    \forkbench for varying \ubsize and $N$}
  \label{plot:fork-copy}
\end{figure}

Figure~\ref{plot:fork-copy} plots FMTC of \forkbench for different
values of \ubsize (64MB and 128MB) and $N$ (2 to 16k) in the baseline
system. As the figure shows, for both values of \ubsize, FMTC
increases with increasing $N$. This is expected as a higher $N$ (more
pages updated by the child) leads to more CoW
operations. However, because of the presence of other read/write
operations (e.g., during the initialization phase of the parent), for
a given value of $N$, FMTC is larger for \ubsize= 64MB compared to
\ubsize= 128MB.  Depending on the value of \ubsize and $N$, anywhere
between 14\% to 66\% of the memory traffic arises from copy
operations. This shows that accelerating copy operations using
RowClone has the potential to significantly improve the performance of
the \fork operation.

\begin{figure}[h]
  \centering
  \includegraphics{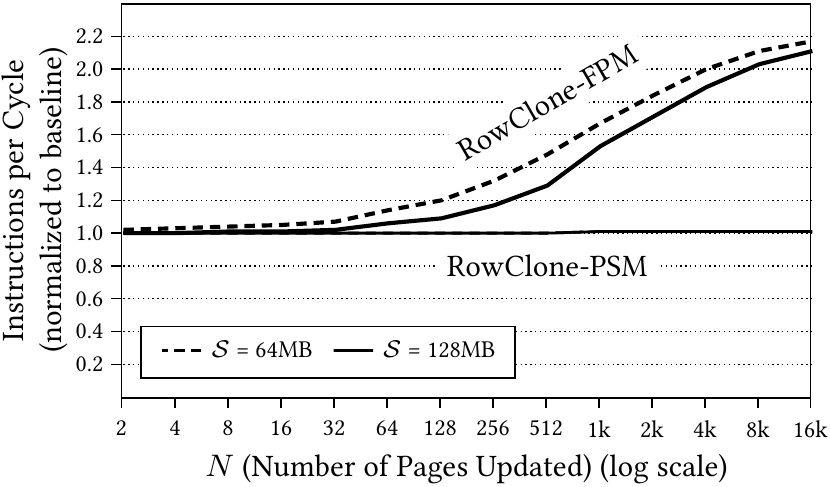}
  \caption[RowClone: \forkbench performance]{Performance
    improvement due to RowClone for \forkbench with different
    values of \ubsize and $N$}
  \label{plot:fork-perf}
\end{figure}

Figure~\ref{plot:fork-perf} plots the performance (IPC) of FPM and
PSM for \forkbench, normalized to that of the baseline system. We
draw two conclusions from the figure. First, FPM improves the
performance of \forkbench for both values of \ubsize and most
values of $N$. The peak performance improvement is 2.2x for $N$ =
16k (30\% on average across all data points). As expected, the
improvement of FPM increases as the number of pages updated
increases. The trend in performance improvement of FPM is similar
to that of FMTC (Figure~\ref{plot:fork-copy}), confirming our
hypothesis that FPM's performance improvement primarily depends on
FMTC. Second, PSM does not provide considerable performance
improvement over the baseline. This is because the large on-chip
cache in the baseline system buffers the writebacks generated by
the copy operations. These writebacks are flushed to memory at a
later point without further delaying the copy operation. As a
result, PSM, which just overlaps the read and write operations
involved in the copy, does not improve latency significantly in
the presence of a large on-chip cache. On the other hand, FPM, by
copying all cache lines from the source row to destination in
parallel, significantly reduces the latency compared to the
baseline (which still needs to read the source blocks from main
memory), resulting in high performance improvement.

\begin{figure}[h]
  \centering
  \includegraphics[scale=0.9]{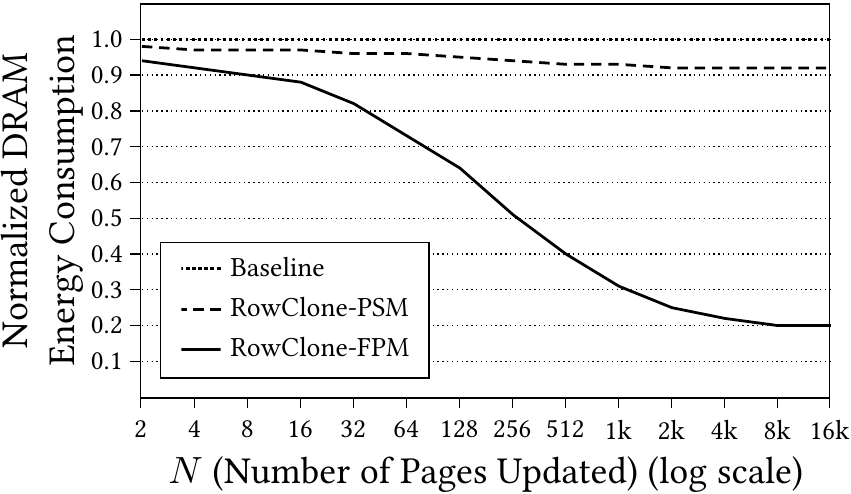}
  \caption[RowClone: Energy consumption for \forkbench]{Comparison
    of DRAM energy consumption of different mechanisms for
    \forkbench (\ubsize = 64MB)}
  \label{plot:fork-energy}
\end{figure}

Figure~\ref{plot:fork-energy} shows the reduction in DRAM energy
consumption (considering both the DRAM and the memory channel) 
of FPM and PSM modes of RowClone compared to that of
the baseline for \forkbench with \ubsize $=64$MB.
Similar to performance, the overall DRAM energy consumption also
depends on the total memory access rate. As a result, RowClone's
potential to reduce DRAM energy depends on the fraction of memory
traffic generated by copy operations. In fact, our results also
show that the DRAM energy reduction due to FPM and PSM correlate
well with FMTC (Figure~\ref{plot:fork-copy}).  By efficiently
performing the copy operations, FPM reduces DRAM energy
consumption by up to 80\% (average 50\%, across all data
points). Similar to FPM, the energy reduction of PSM also
increases with increasing $N$ with a maximum reduction of 9\% for
$N$=16k.

In a system that is agnostic to RowClone, we expect the
performance improvement and energy reduction of RowClone to be in
between that of FPM and PSM. By making the system software aware
of RowClone (Section~\ref{sec:os-changes}),i.e., designing the
system software to be aware of the topology (subarray and bank
organization) of DRAM, as also advocated by various recent
works~\cite{research-mem-sys,ms-sys,row-hammer}, we can
approximate the maximum performance and energy benefits by
increasing the likelihood of the use of FPM.

\subsubsection{Copy/Initialization Intensive Applications}
\label{res:rowclone-ii-apps}

In this section, we analyze the benefits of RowClone on six
copy/initialization intensive applications, including one instance of
the \forkbench described in the previous section. Table~\ref{tab:iias}
describes these applications.

\begin{table}[h]\small
  \centering
  \begin{tabular}{lp{0.87\textwidth}}
  \toprule
  \textbf{Name} & \textbf{Description}\\
  \toprule
  \multirow{1}{*}{\emph{bootup}} & A phase booting up the Debian
  operating system.\\
  \midrule
  \multirow{1}{*}{\emph{compile}} & The compilation phase from the GNU C compiler (while
  running \emph{cc1}).\\
  \midrule
  \multirow{1}{*}{\emph{forkbench}} & An instance of the \forkbench described in
  Section~\ref{sec:res-fork} with \mbox{\ubsize = 64MB} and $N$ = 1k.\\
  \midrule
  \multirow{2}{*}{\emph{mcached}} & Memcached~\cite{memcached}, a memory object
  caching system, a phase inserting many key-value pairs into
  the memcache.\\
  \midrule
  \multirow{1}{*}{\emph{mysql}} & MySQL~\cite{mysql}, an on-disk
  database system, a phase loading the sample \emph{employeedb}\\
  \midrule
  \multirow{2}{*}{\emph{shell}} & A Unix shell script running `find' on a directory
  tree with `ls' on each sub-directory (involves filesystem
  accesses and spawning new processes).\\
  \bottomrule
\end{tabular}

  \caption[RowClone: Copy/initialization-intensive benchmarks]{Copy/Initialization-intensive benchmarks}
  \label{tab:iias}
\end{table}

\begin{figure}[h]
  \centering
  \includegraphics{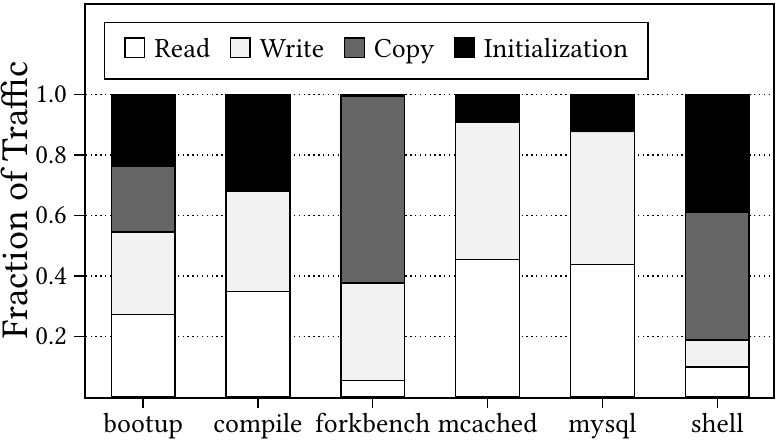}
  \caption[Copy/initialization intensive benchmark: Memory traffic
    breakdown]{Fraction of memory traffic due to read, write, copy
    and initialization}
  \label{plot:memfrac-apps}
\end{figure}

Figure~\ref{plot:memfrac-apps} plots the fraction of memory
traffic due to copy, initialization, and regular read/write
operations for the six applications. For these applications,
between 10\% and 80\% of the memory traffic is generated by copy
and initialization operations.

\begin{figure}[h]
  \centering
  \includegraphics{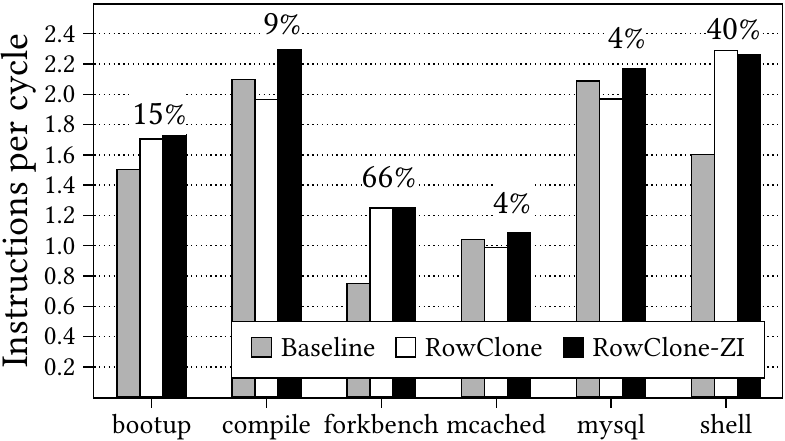}
  \caption[RowClone-ZI performance]{Performance improvement of
    RowClone and RowClone-ZI. \normalfont{Value on top indicates
      percentage improvement of RowClone-ZI over baseline.}}
  \label{plot:perf-apps}
\end{figure}

Figure~\ref{plot:perf-apps} compares the IPC of the baseline with
that of RowClone and a variant of RowClone, RowClone-ZI (described
shortly). The RowClone-based initialization mechanism slightly
degrades performance for the applications which have a negligible
number of copy operations (\emph{mcached}, \emph{compile}, and
\emph{mysql}).

Our further analysis indicated that, for these applications,
although the operating system zeroes out any newly allocated page,
the application typically accesses almost all cache lines of a
page immediately after the page is zeroed out. There are two
phases: 1)~the phase when the OS zeroes out the page, and 2)~the
phase when the application accesses the cache lines of the
page. While the baseline incurs cache misses during phase 1,
RowClone, as a result of performing the zeroing operation
completely in memory, incurs cache misses in phase 2. However, the
baseline zeroing operation is heavily optimized for memory-level
parallelism
(MLP)~\cite{effra,runahead,parbs,blp-prefetch,myin}. Memory-level
parallelism indicates the number of concurrent outstanding misses
to main memory. Higher MLP results in higher overlap in the
latency of the requests. Consequently, higher MLP results in lower
overall latency. In contrast, the cache misses in phase 2 have low
MLP. As a result, incurring the same misses in Phase 2 (as with
RowClone) causes higher overall stall time for the application
(because the latencies for the misses are serialized) than
incurring them in Phase 1 (as in the baseline), resulting in
RowClone's performance degradation compared to the baseline.

To address this problem, RowClone uses a variant called
RowClone-Zero-Insert (RowClone-ZI). RowClone-ZI not only zeroes
out a page in DRAM but it also inserts a zero cache line into the
processor cache corresponding to each cache line in the page that
is zeroed out. By doing so, RowClone-ZI avoids the cache misses
during both phase 1 (zeroing operation) and phase 2 (when the
application accesses the cache lines of the zeroed page).  As a
result, it improves performance for all benchmarks, notably
\forkbench (by 66\%) and \emph{shell} (by 40\%), compared to the
baseline.

Table~\ref{tab:energy-apps} shows the percentage reduction in DRAM
energy and memory bandwidth consumption with RowClone and RowClone-ZI
compared to the baseline. While RowClone significantly reduces both
energy and memory bandwidth consumption for \emph{bootup},
\emph{forkbench} and \emph{shell}, it has negligible impact on both
metrics for the remaining three benchmarks. The lack of energy and
bandwidth benefits in these three applications is due to serial
execution caused by the the cache misses incurred when the processor
accesses the zeroed out pages (i.e., {\em phase 2}, as described
above), which also leads to performance degradation in these workloads
(as also described above). RowClone-ZI, which eliminates the cache
misses in {\em phase 2}, significantly reduces energy consumption
(between 15\% to 69\%) and memory bandwidth consumption (between 16\%
and 81\%) for all benchmarks compared to the baseline. We conclude
that RowClone-ZI can effectively improve performance, memory energy,
and memory bandwidth efficiency in page copy and initialization
intensive single-core workloads.

\begin{table}[h]\small
  \centering
  \begin{tabular}{rrrrr}
  \toprule
  \multirow{2}{*}{\textbf{Application}} &
  \multicolumn{2}{c}{\textbf{Energy Reduction}} &
  \multicolumn{2}{c}{\textbf{Bandwidth Reduction}}\\
  \cmidrule{2-5}
  & \textbf{RowClone} & \textbf{+ZI}
  & \textbf{RowClone} & \textbf{+ZI}\\
  \midrule
  \emph{bootup} & 39\% & 40\% & 49\% & 52\%\vspace{1mm}\\
  \emph{compile} & -2\% & 32\% & 2\% & 47\%\vspace{1mm}\\
  \emph{forkbench} &  69\% & 69\% & 60\% & 60\%\vspace{1mm}\\
  \emph{mcached} & 0\% & 15\% & 0\% & 16\%\vspace{1mm}\\
  \emph{mysql} & -1\% & 17\%  & 0\% & 21\%\vspace{1mm}\\
  \emph{shell} & 68\% & 67\% & 81\% & 81\%\vspace{1mm}\\
  \bottomrule
\end{tabular}


  \caption[RowClone: DRAM energy/bandwidth reduction]{DRAM energy and bandwidth reduction due to RowClone and
  RowClone-ZI (indicated as +ZI)}
  \label{tab:energy-apps}
\end{table}

\subsubsection{Multi-core Evaluations}
\label{sec:multi-core}

As RowClone performs bulk data operations completely within DRAM, it
significantly reduces the memory bandwidth consumed by these
operations. As a result, RowClone can benefit other applications
running concurrently on the same system. We evaluate this benefit of
RowClone by running our copy/initialization-intensive applications
alongside memory-intensive applications from the SPEC CPU2006
benchmark suite~\cite{spec2006} (i.e., those applications with last-level
cache MPKI greater than 1). Table~\ref{tab:benchmarks} lists the set
of applications used for our multi-programmed workloads.

\begin{table}[h]\small
  \centering
  \begin{tabular}{p{0.85\textwidth}}
  \toprule
  \textbf{Copy/Initialization-intensive benchmarks}\vspace{0.5mm}\\
  \emph{bootup},
  \emph{compile}, \emph{forkbench}, \emph{mcached}, \emph{mysql}, \emph{shell}\vspace{1mm}\\
  \toprule
  \textbf{Memory-intensive benchmarks from SPEC CPU2006}\vspace{0.5mm}\\
  \emph{bzip2}, \emph{gcc}, \emph{mcf}, \emph{milc}, \emph{zeusmp},
  \emph{gromacs}, \emph{cactusADM}, \emph{leslie3d}, \emph{namd},
  \emph{gobmk}, \emph{dealII}, \emph{soplex}, \emph{hmmer},
  \emph{sjeng}, \emph{GemsFDTD}, \emph{libquantum}, \emph{h264ref},
  \emph{lbm}, \emph{omnetpp}, \emph{astar}, \emph{wrf},
  \emph{sphinx3}, \emph{xalancbmk}\vspace{1mm}\\
  \bottomrule
\end{tabular}

  \caption[RowClone: Benchmarks for multi-core evaluation]{List of benchmarks used for multi-core evaluation}
  \label{tab:benchmarks}
\end{table}

We generate multi-programmed workloads for 2-core, 4-core and
8-core systems. In each workload, half of the cores run
copy/initialization-intensive benchmarks and the remaining cores
run memory-intensive SPEC benchmarks. Benchmarks from each
category are chosen at random.

\begin{figure}[h]
  \centering
  \includegraphics[scale=0.9]{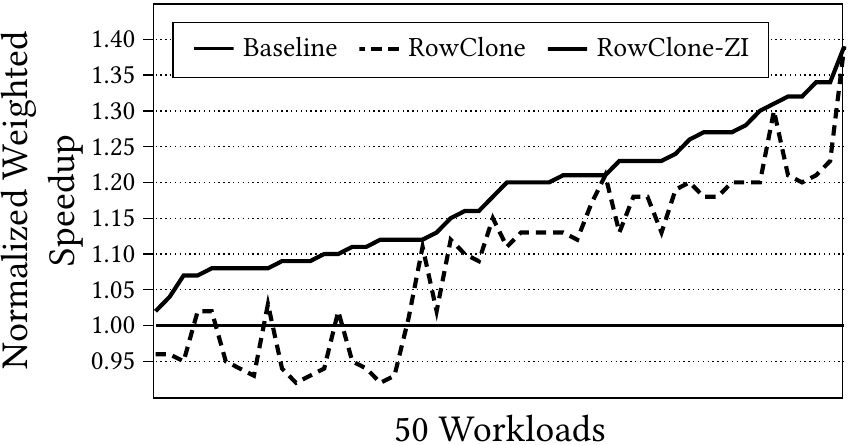}
  \caption[RowClone: 4-core performance]{System performance
    improvement of RowClone for 4-core workloads}
  \label{plot:s-curve}
\end{figure}

Figure~\ref{plot:s-curve} plots the performance improvement due to
RowClone and RowClone-ZI for the 50 4-core workloads we evaluated
(sorted based on the performance improvement due to RowClone-ZI).
Two conclusions are in order. First, although RowClone degrades
performance of certain 4-core workloads (with \emph{compile},
\emph{mcached} or \emph{mysql} benchmarks), it significantly
improves performance for all other workloads (by 10\% across all
workloads). Second, like in our single-core evaluations
(Section~\ref{res:rowclone-ii-apps}), RowClone-ZI eliminates the
performance degradation due to RowClone and consistently
outperforms both the baseline and RowClone for all workloads (20\%
on average).

Table~\ref{tab:multi-core-ws} shows the number of workloads and
six metrics that evaluate the performance, fairness, memory
bandwidth and energy efficiency improvement due to RowClone
compared to the baseline for systems with 2, 4, and 8 cores. We
evaluate fairness using the maximum slowdown metric, which has
been used by many prior works
\cite{app-aware-noc,tcm,atlas,mcp,fst,sms,eaf,asm,bliss,bliss-tpds,mise,fairness-metrics,a2c-noc,firm,dash,cots-mem-inter,aergia,prefetch-aware-srm,a2c-hpca,cots-mem-inter-rtas}
as an indicator of unfairness in the system. Maximum slowdown is
defined as the maximum of the slowdowns of all applications that
are in the multi-core workload.  For all three systems, RowClone
significantly outperforms the baseline on all metrics.

\begin{table}[h]\small
  \centering
  \begin{tabular}{rrrr}
  \toprule
  \textbf{Number of Cores} & \multicolumn{1}{c}{2} & \multicolumn{1}{c}{4} & \multicolumn{1}{c}{8}\\
  \midrule
  \textbf{Number of Workloads} & 138 & 50 & 40\vspace{1mm}\\
  \textbf{Weighted Speedup~\cite{weighted-speedup,weighted-speedup-2} Improvement} & 15\% & 20\% & 27\%\vspace{1mm}\\
  \textbf{Instruction Throughput Improvement} & 14\% & 15\% & 25\%\vspace{1mm}\\
  \textbf{Harmonic Speedup~\cite{hmean} Improvement} & 13\% & 16\% & 29\%\vspace{1mm}\\
  \textbf{Maximum Slowdown~\cite{atlas,tcm,app-aware-noc} Reduction} & 6\% & 12\% & 23\%\vspace{1mm}\\
  \textbf{Memory Bandwidth/Instruction~\cite{fdp} Reduction} & 29\% & 27\% & 28\%\vspace{1mm}\\
  \textbf{Memory Energy/Instruction Reduction} & 19\% & 17\% & 17\%\vspace{1mm}\\
  \bottomrule
\end{tabular}

  \caption[RowClone: Multi-core results]{Multi-core performance, fairness,
  bandwidth, and energy}
  \label{tab:multi-core-ws}
\end{table}

\begin{figure}[h!]
  \centering
  \includegraphics[scale=0.9]{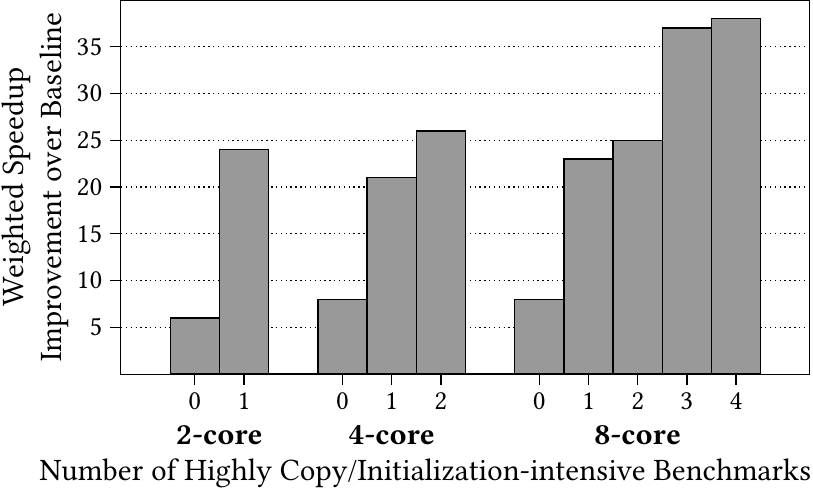}
  \caption[RowClone: Effect of increasing copy/initialization
    intensity]{Effect of increasing copy/initialization intensity}
  \label{plot:multi-trend}
\end{figure}

To provide more insight into the benefits of RowClone on multi-core
systems, we classify our copy/initialization-intensive benchmarks into
two categories: 1) Moderately copy/initialization-intensive
(\emph{compile}, \emph{mcached}, and \emph{mysql}) and highly
copy/initialization-intensive (\emph{bootup}, \emph{forkbench}, and
\emph{shell}). Figure~\ref{plot:multi-trend} shows the average
improvement in weighted speedup for the different multi-core
workloads, categorized based on the number of highly
copy/initialization-intensive benchmarks. As the trends indicate, the
performance improvement increases with increasing number of such
benchmarks for all three multi-core systems, indicating the
effectiveness of RowClone in accelerating bulk copy/initialization
operations.

We conclude that RowClone is an effective mechanism to improve
system performance, energy efficiency and bandwidth efficiency of
future, memory-bandwidth-constrained multi-core systems.

\subsubsection{Memory-Controller-based DMA}
\label{sec:mc-dma}

One alternative way to perform a bulk data operation is to use the
memory controller to complete the operation using the regular DRAM
interface (similar to some prior
approaches~\cite{bulk-copy-initialize,copy-engine}). We refer to
this approach as the memory-controller-based DMA (MC-DMA).  MC-DMA
can potentially avoid the cache pollution caused by inserting
blocks (involved in the copy/initialization) unnecessarily into
the caches. However, it still requires data to be transferred over
the memory bus. Hence, it suffers from the large latency,
bandwidth, and energy consumption associated with the data
transfer. Because the applications used in our evaluations do not
suffer from cache pollution, we expect MC-DMA to perform
comparably or worse than the baseline. In fact, our evaluations
show that MC-DMA degrades performance compared to our baseline by
2\% on average for the six copy/initialization intensive
applications (16\% compared to RowClone). In addition, MC-DMA does
not conserve any DRAM energy, unlike RowClone.

\subsubsection{Other Applications}

\textit{Secure Deallocation.} Most operating systems (e.g.,
Linux~\cite{linux-security}, Windows~\cite{windows-security}, Mac
OS X~\cite{macos-security}) zero out pages newly allocated to a
process. This is done to prevent malicious processes from gaining
access to the data that previously belonged to other processes or
the kernel itself. Not doing so can potentially lead to security
vulnerabilities, as shown by prior
works~\cite{shredding,sunshine,coldboot,disclosure}.

\textit{Process Checkpointing.} Checkpointing is an operation
during which a consistent version of a process state is backed-up,
so that the process can be restored from that state in the
future. This checkpoint-restore primitive is useful in many cases
including high-performance computing servers~\cite{plfs}, software
debugging with reduced overhead~\cite{flashback}, hardware-level
fault and bug tolerance mechanisms~\cite{self-test,hardware-bug,software-bug-tc},
mechanisms to provide consistent updates of persistent memory
state \cite{thynvm}, and speculative OS optimizations to improve
performance~\cite{os-speculation-2,os-speculation-1}. However, to
ensure that the checkpoint is consistent (i.e., the original
process does not update data while the checkpointing is in
progress), the pages of the process are marked with
copy-on-write. As a result, checkpointing often results in a large
number of CoW operations.

\textit{Virtual Machine Cloning/Deduplication.}  Virtual machine
(VM) cloning~\cite{snowflock} is a technique to significantly
reduce the startup cost of VMs in a cloud computing
server. Similarly, deduplication is a technique employed by modern
hypervisors~\cite{esx-server} to reduce the overall memory
capacity requirements of VMs. With this technique, different VMs
share physical pages that contain the same data. Similar to
forking, both these operations likely result in a large number of
CoW operations for pages shared across VMs.

\textit{Page Migration.} Bank conflicts, i.e., concurrent requests to
different rows within the same bank, typically result in reduced row
buffer hit rate and hence degrade both system performance and energy
efficiency. Prior work~\cite{micropages} proposed techniques to
mitigate bank conflicts using page migration. The PSM mode of RowClone
can be used in conjunction with such techniques to 1)~significantly
reduce the migration latency and 2)~make the migrations more
energy-efficient.

\textit{CPU-GPU Communication.} In many current and future
processors, the GPU is or is expected to be integrated on the same
chip with the CPU. Even in such systems where the CPU and GPU
share the same off-chip memory, the off-chip memory is partitioned
between the two devices. As a consequence, whenever a CPU program
wants to offload some computation to the GPU, it has to copy all
the necessary data from the CPU address space to the GPU address
space~\cite{cpu-gpu}. When the GPU computation is finished, all
the data needs to be copied back to the CPU address space. This
copying involves a significant overhead. In fact, a recent work,
Decoupled DMA~\cite{ddma}, motivates this problem and proposes a
solution to mitigate it. By spreading out the GPU address space
over all subarrays and mapping the application data appropriately,
RowClone can significantly speed up these copy operations. Note
that communication between different processors and accelerators
in a heterogeneous System-on-a-chip (SoC) is done similarly to the
CPU-GPU communication and can also be accelerated by RowClone.

\subsection{Applications for IDAO}
\label{sec:idao-eval}

We analyze our mechanism's performance on a real-world bitmap
index library, FastBit~\cite{fastbit}, widely-used in physics
simulations and network analysis. Fastbit can enable faster and
more efficient searching/retrieval compared to B-trees.

To construct an index, FastBit uses multiple bitmap bins, each
corresponding to either a distinct value or a range of
values. FastBit relies on fast bitwise AND/OR operations on these
bitmaps to accelerate \emph{joins} and \emph{range queries}. For
example, to execute a range query, FastBit performs a bitwise OR
of all bitmaps that correspond to the specified range.

We initialized FastBit on our baseline system using the sample
\emph{STAR}~\cite{star} data set. We then ran a set of
indexing-intensive range queries that touch various numbers of
bitmap bins. For each query, we measure the fraction of query
execution time spent on bitwise OR
operations. Table~\ref{table:fraction-or} shows the corresponding
results. For each query, the table shows the number of bitmap bins
involved in the query and the percentage of time spent in bitwise
OR operations. On average, 31\% of the query execution is spent on
bitwise OR operations (with small variance across queries).

\begin{table}[h]
  \caption{Fraction of time spent in OR operations}
  \label{table:fraction-or}
  \centering \begin{tabular}{lccccccc}
  \toprule
  Number of bins & 3 & 9 & 20 & 45 & 98 & 118 & 128\\
  \toprule
  Fraction of time spent in OR & \multirow{1}{*}{29\%} & \multirow{1}{*}{29\%} & \multirow{1}{*}{31\%} & \multirow{1}{*}{32\%} & \multirow{1}{*}{34\%} & \multirow{1}{*}{34\%} & \multirow{1}{*}{34\%}\\
  \bottomrule
\end{tabular}

\end{table}

\begin{figure}[h]
  \centering
  \includegraphics[scale=1.3]{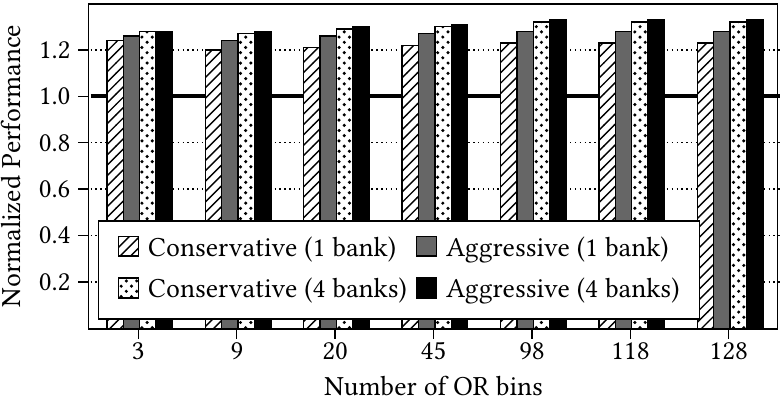}\vspace{-2mm}
  \caption{Range query performance improvement over baseline}\vspace{-2mm}
  \label{plot:fastbit-perf}
\end{figure}

To estimate the performance of our mechanism, we measure the
number of bitwise OR operations required to complete the query. We
then compute the amount of time taken by our mechanism to complete
these operations and then use that to estimate the performance of
the overall query execution. To perform a bitwise OR of more than
two rows, our mechanism is invoked two rows at a
time. Figure~\ref{plot:fastbit-perf} shows the potential
performance improvement using our two mechanisms (conservative and
aggressive), each with either 1 bank or 4 banks.

As our results indicate, our aggressive mechanism with 4 banks
improves the performance of range queries by 30\% (on average)
compared to the baseline, eliminating almost all the overhead of
bitwise operations. As expected, the aggressive mechanism performs
better than the conservative mechanism. Similarly, using more
banks provides better performance. Even if we assume a 2X higher
latency for the triple-row activation, our conservative mechanism
with 1 bank improves performance by 18\% (not shown in the
figure).

\subsection{Recent Works Building on RowClone and IDAO}

Some recent works~\cite{lisa,pinatubo} have built on top of
RowClone and IDAO and have proposed mechanisms to perform bulk
copy and bitwise operations inside memory. As we described in
Section~\ref{sec:bulk-copy}, to copy data across two subarrays in
the same bank, RowClone uses two PSM operations. While this
approach reduces energy consumption compared to existing systems,
it still does not reduce latency. Low-cost Interlinked Sub-Arrays
or LISA~\cite{lisa} addresses this problem by connecting adjacent
subarrays of a bank. LISA exploits the open bitline architecture
and connects the open end of each bitline to the adjacent sense
amplifier using a a transistor. LISA uses these connections to
transfer data more efficiently and quickly across subarrays in the
same bank. Pinatubo~\cite{pinatubo} takes an approach similar to
IDAO and uses Phase Change Memory (PCM) technology to perform
bitwise operations inside a memory chip built using PCM. Pinatubo
enables the PCM sense amplifier to detect fine-grained differences
in cell resistance. With the enhanced sense amplifier, Pinatubo
can perform bitwise AND/OR operations by simultaneously sensing
multiple PCM cells connected to the same sense amplifier.

\section{Conclusion}

In this article, we focused our attention on the problem of data
movement, especially for operations that access a large amount of
data. We first discussed the general notion of Processing in
Memory (PiM) as a potential solution to reducing data movement so
as to achieve better performance and efficiency. PiM adds new
logic structures, sometimes as large as simple processors, near
memory, to perform computation. We then introduced the idea of
Processing using Memory (PuM), which exploits some of the
peripheral structures {\em already existing} inside memory devices
(with minimal changes), to perform other tasks on top of storing
data. PuM is a cost-effective approach as it does not add
significant logic structures near or inside memory.

We developed two new ideas that take the PuM approach and build on
top of DRAM technology. The first idea is RowClone, which exploits
the underlying operation of DRAM to perform bulk copy and
initialization operations completely inside DRAM. RowClone
exploits the fact DRAM cells internally share several buses that
can act as a fast path for copying data across them. The second
idea is In-DRAM AND/OR (IDAO), which exploits the analog operation
of DRAM to perform bulk bitwise AND/OR operations completely
inside DRAM. IDAO exploits the fact that many DRAM cells share the
same sense amplifier and uses simultaneous activation of three
rows of DRAM cells to perform bitwise AND/OR operations
efficiently.

Our evaluations show that both mechanisms (RowClone and IDAO)
improve the performance and energy-efficiency of the respective
operations by more than an order of magnitude. In fact, for
systems that store data in DRAM, these mechanisms are probably as
efficient as any mechanism could be (since they minimize the
amount of data movement). We described many real-world
applications that can exploit RowClone and IDAO, and have
demonstrated significant performance and energy efficiency
improvements using these mechanisms.  Due to its low cost of
implementation and large performance and energy benefits, we
believe Processing using Memory is a very promising and viable
approach to minimize the memory bottleneck in data-intensive
applications. We hope and expect future research will build upon
this approach to demonstrate other techniques that can perform
more operations in memory.

\section*{Acknowledgments}

We thank the members of the SAFARI and LBA research groups, and
the various anonymous reviewers for their valuable feedback on the
multiple works described in this document. We acknowledge the
support of AMD, Google, IBM, Intel, Microsoft, Nvidia, Oracle,
Qualcomm, Samsung, Seagate, and VMWare. This research was
partially supported by NSF (CCF-0953246, CCF-1147397, CCF-1212962,
CNS-1320531), Intel University Research Office Memory Hierarchy
Program, Intel Science and Technology Center for Cloud Computing,
and Semiconductor Research Corporation.

\bibliographystyle{plain}
\bibliography{references}

\begin{thebibliography}{100}

\bibitem{bochs}
{Bochs} {IA-32} emulator project.
\newblock \url{http://bochs.sourceforge.net/}.

\bibitem{fastbit}
{FastBit: An Efficient Compressed Bitmap Index Technology}.
\newblock \url{https://sdm.lbl.gov/fastbit/}.

\bibitem{hbm}
{High Bandwidth Memory DRAM}.
\newblock \url{http://www.jedec.org/standards-documents/docs/jesd235}.

\bibitem{memcached}
{Memcached: A high performance, distributed memory object caching system}.
\newblock \url{http://memcached.org}.

\bibitem{mysql}
{MySQL: An open source database}.
\newblock \url{http://www.mysql.com}.

\bibitem{redis}
Redis - bitmaps.
\newblock \url{http://redis.io/topics/data-types-intro#bitmaps}.

\bibitem{star}
{The STAR experiment}.
\newblock \url{http://www.star.bnl.gov/}.

\bibitem{simics}
{Wind River Simics full system simulation}.
\newblock \url{http://www.windriver.com/products/simics/}.

\bibitem{memsim}
{Memsim}.
\newblock \url{http://safari.ece.cmu.edu/tools.html}, 2012.

\bibitem{ramulator-github}
{Ramulator Source Code}.
\newblock \url{https://github.com/CMU-SAFARI/ramulator}, 2015.

\bibitem{pim-graph}
Junwhan Ahn, Sungpack Hong, Sungjoo Yoo, Onur Mutlu, and Kiyoung Choi.
\newblock {A Scalable Processing-in-memory Accelerator for Parallel Graph
  Processing}.
\newblock In {\em Proceedings of the 42nd Annual International Symposium on
  Computer Architecture}, ISCA '15, pages 105--117, New York, NY, USA, 2015.
  ACM.

\bibitem{pim-enabled-insts}
Junwhan Ahn, Sungjoo Yoo, Onur Mutlu, and Kiyoung Choi.
\newblock {PIM-enabled Instructions: A Low-overhead, Locality-aware
  Processing-in-memory Architecture}.
\newblock In {\em Proceedings of the 42nd Annual International Symposium on
  Computer Architecture}, ISCA '15, pages 336--348, New York, NY, USA, 2015.
  ACM.

\bibitem{data-reorg-3d-stack}
Berkin Akin, Franz Franchetti, and James~C. Hoe.
\newblock {Data Reorganization in Memory Using 3D-stacked DRAM}.
\newblock In {\em Proceedings of the 42nd Annual International Symposium on
  Computer Architecture}, ISCA '15, pages 131--143, New York, NY, USA, 2015.
  ACM.

\bibitem{sms}
Rachata Ausavarungnirun, Kevin Kai-Wei Chang, Lavanya Subramanian, Gabriel~H.
  Loh, and Onur Mutlu.
\newblock {Staged Memory Scheduling: Achieving High Performance and Scalability
  in Heterogeneous Systems}.
\newblock In {\em Proceedings of the 39th Annual International Symposium on
  Computer Architecture}, ISCA '12, pages 416--427, Washington, DC, USA, 2012.
  IEEE Computer Society.

\bibitem{jafar}
Oreoluwa Babarinsa and Stratos Idreos.
\newblock {JAFAR: Near-Data Processing for Databases}.
\newblock In {\em Proceedings of the 2015 ACM SIGMOD International Conference
  on Management of Data}, SIGMOD '15, pages 2069--2070, New York, NY, USA,
  2015. ACM.

\bibitem{plfs}
John Bent, Garth Gibson, Gary Grider, Ben McClelland, Paul Nowoczynski, James
  Nunez, Milo Polte, and Meghan Wingate.
\newblock {PLFS: A Checkpoint Filesystem for Parallel Applications}.
\newblock In {\em Proceedings of the Conference on High Performance Computing
  Networking, Storage and Analysis}, SC '09, pages 21:1--21:12, New York, NY,
  USA, 2009. ACM.

\bibitem{lazypim}
A.~Boroumand, S.~Ghose, B.~Lucia, K.~Hsieh, K.~Malladi, H.~Zheng, and O.~Mutlu.
\newblock {LazyPIM: An Efficient Cache Coherence Mechanism for
  Processing-in-Memory}.
\newblock {\em IEEE Computer Architecture Letters}, PP(99):1--1, 2016.

\bibitem{linux-security}
D.~P. Bovet and M.~Cesati.
\newblock {\em Understanding the Linux Kernel}, page 388.
\newblock O'Reilly Media, 2005.

\bibitem{bmide}
Chee-Yong Chan and Yannis~E. Ioannidis.
\newblock Bitmap index design and evaluation.
\newblock In {\em Proceedings of the 1998 ACM SIGMOD International Conference
  on Management of Data}, SIGMOD '98, pages 355--366, New York, NY, USA, 1998.
  ACM.

\bibitem{os-speculation-2}
Fay Chang and Garth~A. Gibson.
\newblock {Automatic I/O Hint Generation Through Speculative Execution}.
\newblock In {\em Proceedings of the Third Symposium on Operating Systems
  Design and Implementation}, OSDI '99, pages 1--14, Berkeley, CA, USA, 1999.
  USENIX Association.

\bibitem{fly-dram}
Kevin~K. Chang, Abhijith Kashyap, Hasan Hassan, Saugata Ghose, Kevin Hsieh,
  Donghyuk Lee, Tianshi Li, Gennady Pekhimenko, Samira Khan, and Onur Mutlu.
\newblock {Understanding Latency Variation in Modern DRAM Chips: Experimental
  Characterization, Analysis, and Optimization}.
\newblock In {\em Sigmetrics}, 2016.

\bibitem{lisa}
Kevin~K Chang, Prashant~J Nair, Donghyuk Lee, Saugata Ghose, Moinuddin~K
  Qureshi, and Onur Mutlu.
\newblock {Low-Cost Inter-Linked Subarrays (LISA): Enabling Fast Inter-Subarray
  Data Movement in DRAM}.
\newblock In {\em HPCA}, 2016.

\bibitem{hat}
Kevin Kai-Wei Chang, Rachata Ausavarungnirun, Chris Fallin, and Onur Mutlu.
\newblock {HAT: Heterogeneous Adaptive Throttling for On-Chip Networks}.
\newblock In {\em Proceedings of the 2012 IEEE 24th International Symposium on
  Computer Architecture and High Performance Computing}, SBAC-PAD '12, pages
  9--18, Washington, DC, USA, 2012. IEEE Computer Society.

\bibitem{dsarp}
Kevin Kai-Wei Chang, Donghyuk Lee, Zeshan Chishti, Alaa~R Alameldeen, Chris
  Wilkerson, Yoongu Kim, and Onur Mutlu.
\newblock {Improving DRAM performance by parallelizing refreshes with
  accesses}.
\newblock In {\em 2014 IEEE 20th International Symposium on High Performance
  Computer Architecture (HPCA)}, pages 356--367. IEEE, 2014.

\bibitem{shredding}
Jim Chow, Ben Pfaff, Tal Garfinkel, and Mendel Rosenblum.
\newblock {Shredding Your Garbage: Reducing Data Lifetime Through Secure
  Deallocation}.
\newblock In {\em Proceedings of the 14th Conference on USENIX Security
  Symposium - Volume 14}, SSYM'05, pages 22--22, Berkeley, CA, USA, 2005.
  USENIX Association.

\bibitem{hardware-bug}
Kypros Constantinides, Onur Mutlu, and Todd Austin.
\newblock {Online Design Bug Detection: RTL Analysis, Flexible Mechanisms, and
  Evaluation}.
\newblock In {\em Proceedings of the 41st Annual IEEE/ACM International
  Symposium on Microarchitecture}, MICRO 41, pages 282--293, Washington, DC,
  USA, 2008. IEEE Computer Society.

\bibitem{self-test}
Kypros Constantinides, Onur Mutlu, Todd Austin, and Valeria Bertacco.
\newblock {Software-Based Online Detection of Hardware Defects Mechanisms,
  Architectural Support, and Evaluation}.
\newblock In {\em Proceedings of the 40th Annual IEEE/ACM International
  Symposium on Microarchitecture}, MICRO 40, pages 97--108, Washington, DC,
  USA, 2007. IEEE Computer Society.

\bibitem{software-bug-tc}
Kypros Constantinides, Onur Mutlu, Todd Austin, and Valeria Bertacco.
\newblock A flexible software-based framework for online detection of hardware
  defects.
\newblock {\em IEEE Transactions on Computers}, 58(8):1063--1079, 2009.

\bibitem{spec2006}
Standard Performance~Evaluation Corporation.
\newblock {SPEC CPU2006 Benchmark Suite}.
\newblock \url{www.spec.org/cpu2006}, 2006.

\bibitem{bill-dally}
William Dally.
\newblock {GPU Computing to Exascale and Beyond}.
\newblock
  \url{http://www.nvidia.com/content/PDF/sc_2010/theater/Dally_SC10.pdf}.

\bibitem{a2c-noc}
Reetuparna Das, Rachata Ausavarungnirun, Onur Mutlu, Akhilesh Kumar, and Mani
  Azimi.
\newblock {Application-to-core Mapping Policies to Reduce Memory Interference
  in Multi-core Systems}.
\newblock In {\em Proceedings of the 21st International Conference on Parallel
  Architectures and Compilation Techniques}, PACT '12, pages 455--456, New
  York, NY, USA, 2012. ACM.

\bibitem{a2c-hpca}
Reetuparna Das, Rachata Ausavarungnirun, Onur Mutlu, Akhilesh Kumar, and Mani
  Azimi.
\newblock {Application-to-core Mapping Policies to Reduce Memory Interference
  in Multi-core Systems}.
\newblock In {\em HPCA}, 2012.

\bibitem{app-aware-noc}
Reetuparna Das, Onur Mutlu, Thomas Moscibroda, and Chita~R. Das.
\newblock {Application-aware Prioritization Mechanisms for On-chip Networks}.
\newblock In {\em Proceedings of the 42Nd Annual IEEE/ACM International
  Symposium on Microarchitecture}, MICRO 42, pages 280--291, New York, NY, USA,
  2009. ACM.

\bibitem{aergia}
Reetuparna Das, Onur Mutlu, Thomas Moscibroda, and Chita~R. Das.
\newblock {A{\'e}rgia: Exploiting Packet Latency Slack in On-chip Networks}.
\newblock In {\em Proceedings of the 37th Annual International Symposium on
  Computer Architecture}, ISCA '10, pages 106--116, New York, NY, USA, 2010.
  ACM.

\bibitem{diva}
Jeff Draper, Jacqueline Chame, Mary Hall, Craig Steele, Tim Barrett, Jeff
  LaCoss, John Granacki, Jaewook Shin, Chun Chen, Chang~Woo Kang, Ihn Kim, and
  Gokhan Daglikoca.
\newblock {The Architecture of the DIVA Processing-in-memory Chip}.
\newblock In {\em Proceedings of the 16th International Conference on
  Supercomputing}, ICS '02, pages 14--25, New York, NY, USA, 2002. ACM.

\bibitem{sunshine}
Alan~M. Dunn, Michael~Z. Lee, Suman Jana, Sangman Kim, Mark Silberstein,
  Yuanzhong Xu, Vitaly Shmatikov, and Emmett Witchel.
\newblock Eternal sunshine of the spotless machine: Protecting privacy with
  ephemeral channels.
\newblock In {\em Proceedings of the 10th USENIX Conference on Operating
  Systems Design and Implementation}, OSDI'12, pages 61--75, Berkeley, CA, USA,
  2012. USENIX Association.

\bibitem{fst}
Eiman Ebrahimi, Chang~Joo Lee, Onur Mutlu, and Yale~N. Patt.
\newblock {Fairness via Source Throttling: A Configurable and High-performance
  Fairness Substrate for Multi-core Memory Systems}.
\newblock In {\em Proceedings of the Fifteenth Edition of ASPLOS on
  Architectural Support for Programming Languages and Operating Systems},
  ASPLOS XV, pages 335--346, New York, NY, USA, 2010. ACM.

\bibitem{prefetch-aware-srm}
Eiman Ebrahimi, Chang~Joo Lee, Onur Mutlu, and Yale~N. Patt.
\newblock {Prefetch-aware Shared Resource Management for Multi-core Systems}.
\newblock In {\em Proceedings of the 38th Annual International Symposium on
  Computer Architecture}, ISCA '11, pages 141--152, New York, NY, USA, 2011.
  ACM.

\bibitem{multi-pref-mc}
Eiman Ebrahimi, Onur Mutlu, Chang~Joo Lee, and Yale~N. Patt.
\newblock {Coordinated Control of Multiple Prefetchers in Multi-core Systems}.
\newblock In {\em Proceedings of the 42Nd Annual IEEE/ACM International
  Symposium on Microarchitecture}, MICRO 42, pages 316--326, New York, NY, USA,
  2009. ACM.

\bibitem{cram}
Duncan Elliott, Michael Stumm, W.~Martin Snelgrove, Christian Cojocaru, and
  Robert McKenzie.
\newblock {Computational RAM: Implementing Processors in Memory}.
\newblock {\em IEEE Des. Test}, 16(1):32--41, January 1999.

\bibitem{weighted-speedup}
Stijn Eyerman and Lieven Eeckhout.
\newblock {System-Level Performance Metrics for Multiprogram Workloads}.
\newblock {\em IEEE Micro}, 28(3):42--53, May 2008.

\bibitem{nda}
A.~Farmahini-Farahani, Jung~Ho Ahn, K.~Morrow, and Nam~Sung Kim.
\newblock {NDA: Near-DRAM acceleration architecture leveraging commodity DRAM
  devices and standard memory modules}.
\newblock In {\em IEEE 21st International Symposium on High Performance
  Computer Architecture (HPCA), 2015}, pages 283--295, Feb 2015.

\bibitem{programming-flexram}
Basilio~B. Fraguela, Jose Renau, Paul Feautrier, David Padua, and Josep
  Torrellas.
\newblock {Programming the FlexRAM Parallel Intelligent Memory System}.
\newblock In {\em Proceedings of the Ninth ACM SIGPLAN Symposium on Principles
  and Practice of Parallel Programming}, PPoPP '03, pages 49--60, New York, NY,
  USA, 2003. ACM.

\bibitem{pim-analytics}
Mingyu Gao, Grant Ayers, and Christos Kozyrakis.
\newblock {Practical Near-Data Processing for In-Memory Analytics Frameworks}.
\newblock In {\em Proceedings of the 2015 International Conference on Parallel
  Architecture and Compilation (PACT)}, PACT '15, pages 113--124, Washington,
  DC, USA, 2015. IEEE Computer Society.

\bibitem{hrl}
Mingyu Gao and Christos Kozyrakis.
\newblock {HRL: Efficient and Flexible Reconfigurable Logic for Near-Data
  Processing}.
\newblock In {\em HPCA}, 2016.

\bibitem{myin}
Andrew Glew.
\newblock {MLP yes! ILP no}.
\newblock {\em ASPLOS Wild and Crazy Idea Session’98}, 1998.

\bibitem{pim-terasys}
Maya Gokhale, Bill Holmes, and Ken Iobst.
\newblock {Processing in Memory: The Terasys Massively Parallel PIM Array}.
\newblock {\em Computer}, 28(4):23--31, April 1995.

\bibitem{msa3d}
Qi~Guo, Nikolaos Alachiotis, Berkin Akin, Fazle Sadi, Guanglin Xu, Tze~Meng
  Low, Larry Pillegi, James~C. Hoe, and Franz Frachetti.
\newblock {3D-Stacked Memory-Side Acceleration: Accelerator and System Design}.
\newblock In {\em WoNDP}, 2013.

\bibitem{coldboot}
J.~Alex Halderman, Seth~D. Schoen, Nadia Heninger, William Clarkson, William
  Paul, Joseph~A. Calandrino, Ariel~J. Feldman, Jacob Appelbaum, and Edward~W.
  Felten.
\newblock {Lest We Remember: Cold-boot Attacks on Encryption Keys}.
\newblock {\em Commun. ACM}, 52(5):91--98, May 2009.

\bibitem{disclosure}
K.~Harrison and Shouhuai Xu.
\newblock {Protecting Cryptographic Keys from Memory Disclosure Attacks}.
\newblock In {\em 37th Annual IEEE/IFIP International Conference on Dependable
  Systems and Networks}, pages 137--143, June 2007.

\bibitem{emc}
Milad Hashemi, Khubaib, Eiman Ebrahimi, Onur Mutlu, and Yale~N. Patt.
\newblock {Accelerating Dependent Cache Misses with an Enhanced Memory
  Controller}.
\newblock In {\em ISCA}, 2016.

\bibitem{continuous-run-ahead}
Milad Hashemi, Onur Mutlu, and Yale~N. Patt.
\newblock {Continuous Runahead: Transparent Hardware Acceleration for Memory
  Intensive Workloads}.
\newblock In {\em MICRO}, 2016.

\bibitem{chargecache}
Hasan Hassan, Gennady Pekhimenko, Nandita Vijaykumar, Vivek Seshadri, Donghyuk
  Lee, Oguz Ergin, and Onur Mutlu.
\newblock {ChargeCache: Reducing DRAM Latency by Exploiting Row Access
  Locality}.
\newblock In {\em HPCA}, 2016.

\bibitem{ndp-architecture}
Syed~Minhaj Hassan, Sudhakar Yalamanchili, and Saibal Mukhopadhyay.
\newblock {Near Data Processing: Impact and Optimization of 3D Memory System
  Architecture on the Uncore}.
\newblock In {\em Proceedings of the 2015 International Symposium on Memory
  Systems}, MEMSYS '15, pages 11--21, New York, NY, USA, 2015. ACM.

\bibitem{spare-row-mapping}
M.~Horiguchi and K.~Itoh.
\newblock {\em {Nanoscale Memory Repair}}.
\newblock Springer, 2011.

\bibitem{tom}
Kevin Hsieh, Eiman Ebrahimi, Gwangsun Kim, Niladrish Chatterjee, Mike O'Conner,
  Nandita Vijaykumar, Onur Mutlu, and Stephen~W. Keckler.
\newblock {Transparent Offloading and Mapping (TOM): Enabling
  Programmer-Transparent Near-Data Processing in GPU Systems}.
\newblock In {\em ISCA}, 2016.

\bibitem{pica}
Kevin Hsieh, Samira Khan, Nandita Vijaykumar, Kevin~K. Chang, Amirali
  Boroumand, Saugata Ghose, and Onur Mutlu.
\newblock {Accelerating Pointer Chasing in 3D-Stacked Memory: Challenges,
  Mechanisms, Evaluation}.
\newblock In {\em ICCD}, 2016.

\bibitem{s390}
{IBM Corporation}.
\newblock {Enterprise Systems Architecture/390 Principles of Operation}, 2001.

\bibitem{x86-ermsb}
{Intel}.
\newblock {\em {Intel 64 and IA-32 Architectures Optimization Reference
  Manual}}.
\newblock April 2012.

\bibitem{intel-dma}
{Intel}.
\newblock {\em {Intel 64 and IA-32 Architectures Software Developer's Manual}},
  volume~3A, chapter~11, page~12.
\newblock April 2012.

\bibitem{cpu-gpu}
Thomas~B. Jablin, Prakash Prabhu, James~A. Jablin, Nick~P. Johnson, Stephen~R.
  Beard, and David~I. August.
\newblock {Automatic CPU-GPU Communication Management and Optimization}.
\newblock In {\em Proceedings of the 32nd ACM SIGPLAN Conference on Programming
  Language Design and Implementation}, PLDI '11, pages 142--151, New York, NY,
  USA, 2011. ACM.

\bibitem{hmc}
J.~Jeddeloh and B.~Keeth.
\newblock {Hybrid Memory Cube: New DRAM architecture increases density and
  performance}.
\newblock In {\em VLSIT}, pages 87--88, June 2012.

\bibitem{spd}
{JEDEC}.
\newblock {Standard No. 21-C. Annex K: Serial Presence Detect (SPD) for DDR3
  SDRAM Modules}, 2011.

\bibitem{ddr3}
JEDEC.
\newblock {DDR3 SDRAM, JESD79-3F}, 2012.

\bibitem{ddr4}
JEDEC.
\newblock {DDR4 SDRAM Standard}.
\newblock \url{http://www.jedec.org/standards-documents/docs/jesd79-4a}, 2013.

\bibitem{bulk-copy-initialize}
Xiaowei Jiang, Yan Solihin, Li~Zhao, and Ravishankar Iyer.
\newblock {Architecture Support for Improving Bulk Memory Copying and
  Initialization Performance}.
\newblock In {\em PACT}, pages 169--180, Washington, DC, USA, 2009. IEEE
  Computer Society.

\bibitem{sramsod}
Mingu Kang, Min-Sun Keel, Naresh~R Shanbhag, Sean Eilert, and Ken Curewitz.
\newblock {An energy-efficient VLSI architecture for pattern recognition via
  deep embedding of computation in SRAM}.
\newblock In {\em 2014 IEEE International Conference on Acoustics, Speech and
  Signal Processing (ICASSP)}, pages 8326--8330. IEEE, 2014.

\bibitem{flexram}
Yi~Kang, Wei Huang, Seung-Moon Yoo, D.~Keen, Zhenzhou Ge, V.~Lam, P.~Pattnaik,
  and J.~Torrellas.
\newblock {FlexRAM: Toward an Advanced Intelligent Memory System}.
\newblock In {\em Proceedings of the 1999 IEEE International Conference on
  Computer Design}, ICCD '99, pages 192--, Washington, DC, USA, 1999. IEEE
  Computer Society.

\bibitem{dram-cd}
Brent Keeth, R.~Jacob Baker, Brian Johnson, and Feng Lin.
\newblock {\em DRAM Circuit Design: Fundamental and High-Speed Topics}.
\newblock Wiley-IEEE Press, 2nd edition, 2007.

\bibitem{efficacy-error-techniques}
Samira Khan, Donghyuk Lee, Yoongu Kim, Alaa~R. Alameldeen, Chris Wilkerson, and
  Onur Mutlu.
\newblock {The Efficacy of Error Mitigation Techniques for DRAM Retention
  Failures: A Comparative Experimental Study}.
\newblock In {\em The 2014 ACM International Conference on Measurement and
  Modeling of Computer Systems}, SIGMETRICS '14, pages 519--532, New York, NY,
  USA, 2014. ACM.

\bibitem{parbor}
Samira~M. Khan, Donghyuk Lee, and Onur Mutlu.
\newblock {PARBOR: An Efficient System-Level Technique to Detect Data-Dependent
  Failures in DRAM}.
\newblock In {\em DSN}, 2016.

\bibitem{cots-mem-inter-rtas}
Hyoseung Kim, Dionisio de~Niz, Bj{\"o}rn Andersson, Mark Klein, Onur Mutlu, and
  Ragunathan Rajkumar.
\newblock {Bounding and reducing memory interference in COTS-based multi-core
  systems}.
\newblock In {\em RTAS}, 2014.

\bibitem{cots-mem-inter}
Hyoseung Kim, Dionisio de~Niz, Bj{\"o}rn Andersson, Mark Klein, Onur Mutlu, and
  Ragunathan Rajkumar.
\newblock {Bounding and reducing memory interference in COTS-based multi-core
  systems}.
\newblock {\em Real-Time Systems}, 52(3):356--395, 2016.

\bibitem{row-hammer}
Yoongu Kim, Ross Daly, Jeremie Kim, Chris Fallin, Ji~Hye Lee, Donghyuk Lee,
  Chris Wilkerson, Konrad Lai, and Onur Mutlu.
\newblock {Flipping Bits in Memory Without Accessing Them: An Experimental
  Study of DRAM Disturbance Errors}.
\newblock In {\em Proceeding of the 41st Annual International Symposium on
  Computer Architecuture}, ISCA '14, pages 361--372, Piscataway, NJ, USA, 2014.
  IEEE Press.

\bibitem{atlas}
Yoongu Kim, Dongsu Han, O.~Mutlu, and M.~Harchol-Balter.
\newblock {ATLAS: A scalable and high-performance scheduling algorithm for
  multiple memory controllers}.
\newblock In {\em IEEE 16th International Symposium on High Performance
  Computer Architecture}, pages 1--12, Jan 2010.

\bibitem{tcm}
Yoongu Kim, Michael Papamichael, Onur Mutlu, and Mor Harchol-Balter.
\newblock {Thread Cluster Memory Scheduling: Exploiting Differences in Memory
  Access Behavior}.
\newblock In {\em Proceedings of the 2010 43rd Annual IEEE/ACM International
  Symposium on Microarchitecture}, MICRO '43, pages 65--76, Washington, DC,
  USA, 2010. IEEE Computer Society.

\bibitem{salp}
Yoongu Kim, Vivek Seshadri, Donghyuk Lee, Jamie Liu, and Onur Mutlu.
\newblock {A Case for Exploiting Subarray-level Parallelism (SALP) in DRAM}.
\newblock In {\em Proceedings of the 39th Annual International Symposium on
  Computer Architecture}, ISCA '12, pages 368--379, Washington, DC, USA, 2012.
  IEEE Computer Society.

\bibitem{ramulator}
Yoongu Kim, Weikun Yang, and Onur Mutlu.
\newblock {Ramulator: A Fast and Extensible DRAM Simulator}.
\newblock {\em IEEE Comput. Archit. Lett.}, 15(1):45--49, January 2016.

\bibitem{execube}
Peter~M. Kogge.
\newblock {EXECUBE: A New Architecture for Scaleable MPPs}.
\newblock In {\em ICPP}, pages 77--84, Washington, DC, USA, 1994. IEEE Computer
  Society.

\bibitem{snowflock}
Horacio~Andr{\'e}s Lagar-Cavilla, Joseph~Andrew Whitney, Adin~Matthew Scannell,
  Philip Patchin, Stephen~M. Rumble, Eyal de~Lara, Michael Brudno, and Mahadev
  Satyanarayanan.
\newblock {SnowFlock: Rapid Virtual Machine Cloning for Cloud Computing}.
\newblock In {\em Proceedings of the 4th ACM European Conference on Computer
  Systems}, EuroSys '09, pages 1--12, New York, NY, USA, 2009. ACM.

\bibitem{pcm1}
Benjamin~C. Lee, Engin Ipek, Onur Mutlu, and Doug Burger.
\newblock {Architecting Phase Change Memory As a Scalable DRAM Alternative}.
\newblock In {\em Proceedings of the 36th Annual International Symposium on
  Computer Architecture}, ISCA '09, pages 2--13, New York, NY, USA, 2009. ACM.

\bibitem{pcm-scalable}
Benjamin~C. Lee, Engin Ipek, Onur Mutlu, and Doug Burger.
\newblock {Phase Change Memory Architecture and the Quest for Scalability}.
\newblock {\em Commun. ACM}, 53(7):99--106, July 2010.

\bibitem{pcm5}
Benjamin~C. Lee, Ping Zhou, Jun Yang, Youtao Zhang, Bo~Zhao, Engin Ipek, Onur
  Mutlu, and Doug Burger.
\newblock {Phase-Change Technology and the Future of Main Memory}.
\newblock {\em IEEE Micro}, 30(1):143--143, January 2010.

\bibitem{blp-prefetch}
Chang~Joo Lee, Veynu Narasiman, Onur Mutlu, and Yale~N. Patt.
\newblock Improving memory bank-level parallelism in the presence of
  prefetching.
\newblock In {\em Proceedings of the 42nd Annual IEEE/ACM International
  Symposium on Microarchitecture}, MICRO 42, pages 327--336, New York, NY, USA,
  2009. ACM.

\bibitem{smla}
Donghyuk Lee, Saugata Ghose, Gennady Pekhimenko, Samira Khan, and Onur Mutlu.
\newblock {Simultaneous Multi-Layer Access: Improving 3D-Stacked Memory
  Bandwidth at Low Cost}.
\newblock {\em ACM Trans. Archit. Code Optim.}, 12(4):63:1--63:29, January
  2016.

\bibitem{al-dram}
Donghyuk Lee, Yoongu Kim, Gennady Pekhimenko, Samira~Manabi Khan, Vivek
  Seshadri, Kevin Kai-Wei Chang, and Onur Mutlu.
\newblock {{Adaptive-latency DRAM: Optimizing DRAM timing for the
  common-case}}.
\newblock In {\em {HPCA}}, pages 489--501. {IEEE}, 2015.

\bibitem{tl-dram}
Donghyuk Lee, Yoongu Kim, Vivek Seshadri, Jamie Liu, Lavanya Subramanian, and
  Onur Mutlu.
\newblock {Tiered-latency DRAM: A Low Latency and Low Cost DRAM Architecture}.
\newblock In {\em Proceedings of the 2013 IEEE 19th International Symposium on
  High Performance Computer Architecture (HPCA)}, HPCA '13, pages 615--626,
  Washington, DC, USA, 2013. IEEE Computer Society.

\bibitem{ddma}
Donghyuk Lee, Lavanya Subramanian, Rachata Ausavarungnirun, Jongmoo Choi, and
  Onur Mutlu.
\newblock {Decoupled Direct Memory Access: Isolating CPU and IO Traffic by
  Leveraging a Dual-Data-Port DRAM}.
\newblock In {\em Proceedings of the 2015 International Conference on Parallel
  Architecture and Compilation (PACT)}, PACT '15, pages 174--187, Washington,
  DC, USA, 2015. IEEE Computer Society.

\bibitem{pinatubo}
Shuangchen Li, Cong Xu, Qiaosha Zou, Jishen Zhao, Yu~Lu, and Yuan Xie.
\newblock {Pinatubo: A Processing-in-Memory Architecture for Bulk Bitwise
  Operations in Emerging Non-Volatile Memories}.
\newblock In {\em Proceedings of the 53rd Annual Design Automation Conference},
  page 173. ACM, 2016.

\bibitem{data-retention}
Jamie Liu, Ben Jaiyen, Yoongu Kim, Chris Wilkerson, and Onur Mutlu.
\newblock {An Experimental Study of Data Retention Behavior in Modern DRAM
  Devices: Implications for Retention Time Profiling Mechanisms}.
\newblock In {\em Proceedings of the 40th Annual International Symposium on
  Computer Architecture}, ISCA '13, pages 60--71, New York, NY, USA, 2013. ACM.

\bibitem{raidr}
Jamie Liu, Ben Jaiyen, Richard Veras, and Onur Mutlu.
\newblock {RAIDR: Retention-Aware Intelligent DRAM Refresh}.
\newblock In {\em Proceedings of the 39th Annual International Symposium on
  Computer Architecture}, ISCA '12, pages 1--12, Washington, DC, USA, 2012.
  IEEE Computer Society.

\bibitem{3d-stacking}
Gabriel~H. Loh.
\newblock {3D-Stacked Memory Architectures for Multi-core Processors}.
\newblock In {\em Proceedings of the 35th Annual International Symposium on
  Computer Architecture}, ISCA '08, pages 453--464, Washington, DC, USA, 2008.
  IEEE Computer Society.

\bibitem{hmean}
Kun Luo, J.~Gummaraju, and M.~Franklin.
\newblock {Balancing thoughput and fairness in SMT processors}.
\newblock In {\em Performance Analysis of Systems and Software, 2001. ISPASS.
  2001 IEEE International Symposium on}, pages 164--171, 2001.

\bibitem{row-buffer-nvm}
Justin Meza, Jing Li, and Onur Mutlu.
\newblock {A Case for Small Row Buffers in Non-volatile Main Memories}.
\newblock In {\em Proceedings of the 2012 IEEE 30th International Conference on
  Computer Design (ICCD 2012)}, ICCD '12, pages 484--485, Washington, DC, USA,
  2012. IEEE Computer Society.

\bibitem{gp-simd}
Amir Morad, Leonid Yavits, and Ran Ginosar.
\newblock {GP-SIMD Processing-in-Memory}.
\newblock {\em ACM Trans. Archit. Code Optim.}, 11(4):53:1--53:26, January
  2015.

\bibitem{mcp}
Sai~Prashanth Muralidhara, Lavanya Subramanian, Onur Mutlu, Mahmut Kandemir,
  and Thomas Moscibroda.
\newblock {Reducing Memory Interference in Multicore Systems via
  Application-aware Memory Channel Partitioning}.
\newblock In {\em Proceedings of the 44th Annual IEEE/ACM International
  Symposium on Microarchitecture}, MICRO-44, pages 374--385, New York, NY, USA,
  2011. ACM.

\bibitem{effra}
Onur Mutlu.
\newblock {\em {Efficient Runahead Execution Processors}}.
\newblock PhD thesis, Austin, TX, USA, 2006.
\newblock AAI3263366.

\bibitem{ms-sys}
Onur Mutlu.
\newblock {Memory Scaling: A Systems Architecture Perspective}.
\newblock In {\em IMW}, 2014.

\bibitem{parbs}
Onur Mutlu and Thomas Moscibroda.
\newblock {Parallelism-Aware Batch Scheduling: Enhancing Both Performance and
  Fairness of Shared DRAM Systems}.
\newblock In {\em Proceedings of the 35th Annual International Symposium on
  Computer Architecture}, ISCA '08, pages 63--74, Washington, DC, USA, 2008.
  IEEE Computer Society.

\bibitem{runahead}
Onur Mutlu, Jared Stark, Chris Wilkerson, and Yale~N. Patt.
\newblock {Runahead Execution: An Alternative to Very Large Instruction Windows
  for Out-of-Order Processors}.
\newblock In {\em Proceedings of the 9th International Symposium on
  High-Performance Computer Architecture}, HPCA '03, pages 129--, Washington,
  DC, USA, 2003. IEEE Computer Society.

\bibitem{research-mem-sys}
Onur Mutlu and Lavanya Subramanian.
\newblock {Research Problems and Opportunities in Memory Systems}.
\newblock {\em SuperFRI}, 2014.

\bibitem{bmidc}
Elizabeth O'Neil, Patrick O'Neil, and Kesheng Wu.
\newblock {Bitmap Index Design Choices and Their Performance Implications}.
\newblock In {\em Proceedings of the 11th International Database Engineering
  and Applications Symposium}, IDEAS '07, pages 72--84, Washington, DC, USA,
  2007. IEEE Computer Society.

\bibitem{active-pages}
Mark Oskin, Frederic~T. Chong, and Timothy Sherwood.
\newblock {Active Pages: A Computation Model for Intelligent Memory}.
\newblock In {\em Proceedings of the 25th Annual International Symposium on
  Computer Architecture}, ISCA '98, pages 192--203, Washington, DC, USA, 1998.
  IEEE Computer Society.

\bibitem{gpu-pim}
Ashutosh Patnaik, Xulong Tang, Adwait Jog, Onur Kayiran, Asit~K. Mishra,
  Mahmut~T. Kandemir, Onur Mutlu, and Chita~R. Das.
\newblock {Scheduling Techniques for GPU Architectures with
  Processing-In-Memory Capabilities}.
\newblock In {\em PACT}, 2016.

\bibitem{iram}
David Patterson, Thomas Anderson, Neal Cardwell, Richard Fromm, Kimberly
  Keeton, Christoforos Kozyrakis, Randi Thomas, and Katherine Yelick.
\newblock {A Case for Intelligent RAM}.
\newblock {\em IEEE Micro}, 17(2):34--44, March 1997.

\bibitem{avatar}
M.~K. Qureshi, D.~H. Kim, S.~Khan, P.~J. Nair, and O.~Mutlu.
\newblock {AVATAR: A Variable-Retention-Time (VRT) Aware Refresh for DRAM
  Systems}.
\newblock In {\em 2015 45th Annual IEEE/IFIP International Conference on
  Dependable Systems and Networks}, pages 427--437, June 2015.

\bibitem{pcm2}
Moinuddin~K. Qureshi, Vijayalakshmi Srinivasan, and Jude~A. Rivers.
\newblock {Scalable High Performance Main Memory System Using Phase-change
  Memory Technology}.
\newblock In {\em Proceedings of the 36th Annual International Symposium on
  Computer Architecture}, ISCA '09, pages 24--33, New York, NY, USA, 2009. ACM.

\bibitem{rambus-power}
Rambus.
\newblock {DRAM power model}, 2010.

\bibitem{pcm-ibm}
S.~Raoux, G.~W. Burr, M.~J. Breitwisch, C.~T. Rettner, Y.-C. Chen, R.~M.
  Shelby, M.~Salinga, D.~Krebs, S.-H. Chen, H.-L. Lung, and C.~H. Lam.
\newblock {Phase-change Random Access Memory: A Scalable Technology}.
\newblock {\em IBM J. Res. Dev.}, 52(4):465--479, July 2008.

\bibitem{thynvm}
Jinglei Ren, Jishen Zhao, Samira Khan, Jongmoo Choi, Yongwei Wu, and Onur
  Mutlu.
\newblock {ThyNVM: Enabling Software-transparent Crash Consistency in
  Persistent Memory Systems}.
\newblock In {\em Proceedings of the 48th International Symposium on
  Microarchitecture}, MICRO-48, pages 672--685, New York, NY, USA, 2015. ACM.

\bibitem{windows-security}
M.~E. Russinovich, D.~A. Solomon, and A.~Ionescu.
\newblock {\em {Windows Internals}}, page 701.
\newblock Microsoft Press, 2009.

\bibitem{fork-exp}
R.~F. Sauers, C.~P. Ruemmler, and P.~S. Weygant.
\newblock {\em {HP-UX 11i Tuning and Performance}}, chapter 8. Memory
  Bottlenecks.
\newblock Prentice Hall, 2004.

\bibitem{dbi}
Vivek Seshadri, Abhishek Bhowmick, Onur Mutlu, Phillip~B. Gibbons, Michael~A.
  Kozuch, and Todd~C. Mowry.
\newblock {The Dirty-Block Index}.
\newblock In {\em Proceeding of the 41st Annual International Symposium on
  Computer Architecuture}, ISCA '14, pages 157--168, Piscataway, NJ, USA, 2014.
  IEEE Press.

\bibitem{bitwise-cal}
Vivek Seshadri, Kevin Hsieh, Amirali Boroumand, Donghyuk Lee, Michael~A.
  Kozuch, Onur Mutlu, Phillip~B. Gibbons, and Todd~C. Mowry.
\newblock {Fast Bulk Bitwise AND and OR in DRAM}.
\newblock {\em IEEE Comput. Archit. Lett.}, 14(2):127--131, July 2015.

\bibitem{rowclone}
Vivek Seshadri, Yoongu Kim, Chris Fallin, Donghyuk Lee, Rachata
  Ausavarungnirun, Gennady Pekhimenko, Yixin Luo, Onur Mutlu, Phillip~B.
  Gibbons, Michael~A. Kozuch, and Todd~C. Mowry.
\newblock {RowClone: Fast and Energy-efficient in-DRAM Bulk Data Copy and
  Initialization}.
\newblock In {\em Proceedings of the 46th Annual IEEE/ACM International
  Symposium on Microarchitecture}, MICRO-46, pages 185--197, New York, NY, USA,
  2013. ACM.

\bibitem{gsdram}
Vivek Seshadri, Thomas Mullins, Amirali Boroumand, Onur Mutlu, Phillip~B.
  Gibbons, Michael~A. Kozuch, and Todd~C. Mowry.
\newblock {Gather-Scatter DRAM: In-DRAM Address Translation to Improve the
  Spatial Locality of Non-unit Strided Accesses}.
\newblock In {\em Proceedings of the 48th International Symposium on
  Microarchitecture}, MICRO-48, pages 267--280, New York, NY, USA, 2015. ACM.

\bibitem{eaf}
Vivek Seshadri, Onur Mutlu, Michael~A. Kozuch, and Todd~C. Mowry.
\newblock {The Evicted-address Filter: A Unified Mechanism to Address Both
  Cache Pollution and Thrashing}.
\newblock In {\em Proceedings of the 21st International Conference on Parallel
  Architectures and Compilation Techniques}, PACT '12, pages 355--366, New
  York, NY, USA, 2012. ACM.

\bibitem{icp}
Vivek Seshadri, Samihan Yedkar, Hongyi Xin, Onur Mutlu, Phillip~B. Gibbons,
  Michael~A. Kozuch, and Todd~C. Mowry.
\newblock {Mitigating Prefetcher-Caused Pollution Using Informed Caching
  Policies for Prefetched Blocks}.
\newblock {\em ACM Trans. Archit. Code Optim.}, 11(4):51:1--51:22, January
  2015.

\bibitem{isaac}
Ali Shafiee, Anirban Nag, Naveen Muralimanohar, Rajeev Balasubramonian,
  John~Paul Strachan, Miao Hu, R~Stanley Williams, and Vivek Srikumar.
\newblock {ISAAC: A Convolutional Neural Network Accelerator with In-Situ
  Analog Arithmetic in Crossbars}.
\newblock In {\em Proc. ISCA}, 2016.

\bibitem{non-von-machine}
David~Elliot Shaw, Salvatore Stolfo, Hussein Ibrahim, Bruce~K. Hillyer, Jim
  Andrews, and Gio Wiederhold.
\newblock {The NON-VON Database Machine: An Overview}.
\newblock \url{http://hdl.handle.net/10022/AC:P:11530.}, 1981.

\bibitem{macos-security}
A.~Singh.
\newblock {\em {Mac OS X Internals: A Systems Approach}}.
\newblock Addison-Wesley Professional, 2006.

\bibitem{weighted-speedup-2}
Allan Snavely and Dean~M. Tullsen.
\newblock {Symbiotic Jobscheduling for a Simultaneous Multithreaded Processor}.
\newblock In {\em Proceedings of the Ninth International Conference on
  Architectural Support for Programming Languages and Operating Systems},
  ASPLOS IX, pages 234--244, New York, NY, USA, 2000. ACM.

\bibitem{fdp}
Santhosh Srinath, Onur Mutlu, Hyesoon Kim, and Yale~N. Patt.
\newblock {Feedback Directed Prefetching: Improving the Performance and
  Bandwidth-Efficiency of Hardware Prefetchers}.
\newblock In {\em Proceedings of the 2007 IEEE 13th International Symposium on
  High Performance Computer Architecture}, HPCA '07, pages 63--74, Washington,
  DC, USA, 2007. IEEE Computer Society.

\bibitem{flashback}
Sudarshan~M. Srinivasan, Srikanth Kandula, Christopher~R. Andrews, and Yuanyuan
  Zhou.
\newblock {Flashback: A Lightweight Extension for Rollback and Deterministic
  Replay for Software Debugging}.
\newblock In {\em Proceedings of the Annual Conference on USENIX Annual
  Technical Conference}, ATEC '04, pages 3--3, Berkeley, CA, USA, 2004. USENIX
  Association.

\bibitem{lim-computer}
Harold~S. Stone.
\newblock {A Logic-in-Memory Computer}.
\newblock {\em IEEE Trans. Comput.}, 19(1):73--78, January 1970.

\bibitem{bliss}
L.~Subramanian, D.~Lee, V.~Seshadri, H.~Rastogi, and O.~Mutlu.
\newblock {The Blacklisting Memory Scheduler: Achieving high performance and
  fairness at low cost}.
\newblock In {\em ICCD}, 2014.

\bibitem{bliss-tpds}
L.~Subramanian, D.~Lee, V.~Seshadri, H.~Rastogi, and O.~Mutlu.
\newblock {BLISS: Balancing Performance, Fairness and Complexity in Memory
  Access Schedyuling}.
\newblock {\em IEEE Transactions on Parallel and Distributed Systems}, 2016.

\bibitem{mise}
L.~Subramanian, V.~Seshadri, Yoongu Kim, B.~Jaiyen, and O.~Mutlu.
\newblock {MISE: Providing performance predictability and improving fairness in
  shared main memory systems}.
\newblock In {\em IEEE 19th International Symposium on High Performance
  Computer Architecture}, pages 639--650, Feb 2013.

\bibitem{asm}
Lavanya Subramanian, Vivek Seshadri, Arnab Ghosh, Samira Khan, and Onur Mutlu.
\newblock {The Application Slowdown Model: Quantifying and Controlling the
  Impact of Inter-application Interference at Shared Caches and Main Memory}.
\newblock In {\em Proceedings of the 48th International Symposium on
  Microarchitecture}, MICRO-48, pages 62--75, New York, NY, USA, 2015. ACM.

\bibitem{micropages}
Kshitij Sudan, Niladrish Chatterjee, David Nellans, Manu Awasthi, Rajeev
  Balasubramonian, and Al~Davis.
\newblock {Micro-pages: Increasing DRAM Efficiency with Locality-aware Data
  Placement}.
\newblock In {\em Proceedings of the Fifteenth Edition of ASPLOS on
  Architectural Support for Programming Languages and Operating Systems},
  ASPLOS XV, pages 219--230, New York, NY, USA, 2010. ACM.

\bibitem{data-access-opt-pim}
Zehra Sura, Arpith Jacob, Tong Chen, Bryan Rosenburg, Olivier Sallenave, Carlo
  Bertolli, Samuel Antao, Jose Brunheroto, Yoonho Park, Kevin O'Brien, and Ravi
  Nair.
\newblock Data access optimization in a processing-in-memory system.
\newblock In {\em Proceedings of the 12th ACM International Conference on
  Computing Frontiers}, CF '15, pages 6:1--6:8, New York, NY, USA, 2015. ACM.

\bibitem{rethinking-dram}
Aniruddha~N. Udipi, Naveen Muralimanohar, Niladrish Chatterjee, Rajeev
  Balasubramonian, Al~Davis, and Norman~P. Jouppi.
\newblock {Rethinking DRAM Design and Organization for Energy-constrained
  Multi-cores}.
\newblock In {\em Proceedings of the 37th Annual International Symposium on
  Computer Architecture}, ISCA '10, pages 175--186, New York, NY, USA, 2010.
  ACM.

\bibitem{dash}
Hiroyuki Usui, Lavanya Subramanian, Kevin Kai-Wei Chang, and Onur Mutlu.
\newblock {DASH: Deadline-Aware High-Performance Memory Scheduler for
  Heterogeneous Systems with Hardware Accelerators}.
\newblock {\em ACM Trans. Archit. Code Optim.}, 12(4):65:1--65:28, January
  2016.

\bibitem{fairness-metrics}
Hans Vandierendonck and Andre Seznec.
\newblock {Fairness Metrics for Multi-Threaded Processors}.
\newblock {\em IEEE Comput. Archit. Lett.}, 10(1):4--7, January 2011.

\bibitem{esx-server}
Carl~A. Waldspurger.
\newblock {Memory Resource Management in VMware ESX Server}.
\newblock {\em SIGOPS Oper. Syst. Rev.}, 36(SI):181--194, December 2002.

\bibitem{threaded-module}
F.A. Ware and C.~Hampel.
\newblock {Improving Power and Data Efficiency with Threaded Memory Modules}.
\newblock In {\em ICCD}, 2006.

\bibitem{os-speculation-1}
Benjamin Wester, Peter~M. Chen, and Jason Flinn.
\newblock {Operating System Support for Application-specific Speculation}.
\newblock In {\em Proceedings of the Sixth Conference on Computer Systems},
  EuroSys '11, pages 229--242, New York, NY, USA, 2011. ACM.

\bibitem{pcm3}
H.~S.~P. Wong, S.~Raoux, S.~Kim, J.~Liang, J.~P. Reifenberg, B.~Rajendran,
  M.~Asheghi, and K.~E. Goodson.
\newblock {Phase Change Memory}.
\newblock {\em Proceedings of the IEEE}, 98(12):2201--2227, Dec 2010.

\bibitem{bicompression}
Kesheng Wu, Ekow~J. Otoo, and Arie Shoshani.
\newblock {Compressing Bitmap Indexes for Faster Search Operations}.
\newblock In {\em Proceedings of the 14th International Conference on
  Scientific and Statistical Database Management}, SSDBM '02, pages 99--108,
  Washington, DC, USA, 2002. IEEE Computer Society.

\bibitem{why-nothing-matters}
Xi~Yang, Stephen~M. Blackburn, Daniel Frampton, Jennifer~B. Sartor, and
  Kathryn~S. McKinley.
\newblock {Why Nothing Matters: The Impact of Zeroing}.
\newblock In {\em OOPSLA}, pages 307--324, New York, NY, USA, 2011. ACM.

\bibitem{rbla}
HanBin Yoon, Justin Meza, Rachata Ausavarungnirun, Rachel Harding, and Onur
  Mutlu.
\newblock {Row Buffer Locality Aware Caching Policies for Hybrid Memories}.
\newblock In {\em Proceedings of the 2012 IEEE 30th International Conference on
  Computer Design (ICCD 2012)}, ICCD '12, pages 337--344, Washington, DC, USA,
  2012. IEEE Computer Society.

\bibitem{mlc-pcm}
Hanbin Yoon, Justin Meza, Naveen Muralimanohar, Norman~P. Jouppi, and Onur
  Mutlu.
\newblock {Efficient Data Mapping and Buffering Techniques for Multilevel Cell
  Phase-Change Memories}.
\newblock {\em ACM Trans. Archit. Code Optim.}, 11(4):40:1--40:25, December
  2014.

\bibitem{top-pim}
Dongping Zhang, Nuwan Jayasena, Alexander Lyashevsky, Joseph~L. Greathouse,
  Lifan Xu, and Michael Ignatowski.
\newblock {TOP-PIM: Throughput-oriented Programmable Processing in Memory}.
\newblock In {\em Proceedings of the 23rd International Symposium on
  High-performance Parallel and Distributed Computing}, HPDC '14, pages 85--98,
  New York, NY, USA, 2014. ACM.

\bibitem{half-dram}
Tao Zhang, Ke~Chen, Cong Xu, Guangyu Sun, Tao Wang, and Yuan Xie.
\newblock {Half-DRAM: A High-bandwidth and Low-power DRAM Architecture from the
  Rethinking of Fine-grained Activation}.
\newblock In {\em Proceeding of the 41st Annual International Symposium on
  Computer Architecuture}, ISCA '14, pages 349--360, Piscataway, NJ, USA, 2014.
  IEEE Press.

\bibitem{firm}
Jishen Zhao, Onur Mutlu, and Yuan Xie.
\newblock {FIRM: Fair and High-Performance Memory Control for Persistent Memory
  Systems}.
\newblock In {\em Proceedings of the 47th Annual IEEE/ACM International
  Symposium on Microarchitecture}, MICRO-47, pages 153--165, Washington, DC,
  USA, 2014. IEEE Computer Society.

\bibitem{copy-engine}
Li~Zhao, Laxmi~N. Bhuyan, Ravi Iyer, Srihari Makineni, and Donald Newell.
\newblock {Hardware Support for Accelerating Data Movement in Server Platform}.
\newblock {\em IEEE Trans. Comput.}, 56(6):740--753, June 2007.

\bibitem{mini-rank}
Hongzhong Zheng, Jiang Lin, Zhao Zhang, Eugene Gorbatov, Howard David, and
  Zhichun Zhu.
\newblock {Mini-rank: Adaptive DRAM Architecture for Improving Memory Power
  Efficiency}.
\newblock In {\em MICRO}, 2008.

\bibitem{pcm4}
Ping Zhou, Bo~Zhao, Jun Yang, and Youtao Zhang.
\newblock {A Durable and Energy Efficient Main Memory Using Phase Change Memory
  Technology}.
\newblock In {\em Proceedings of the 36th Annual International Symposium on
  Computer Architecture}, ISCA '09, pages 14--23, New York, NY, USA, 2009. ACM.

\bibitem{spmm-mul-lim}
Q.~Zhu, T.~Graf, H.~E. Sumbul, L.~Pileggi, and F.~Franchetti.
\newblock {Accelerating sparse matrix-matrix multiplication with 3D-stacked
  logic-in-memory hardware}.
\newblock In {\em High Performance Extreme Computing Conference (HPEC), 2013
  IEEE}, pages 1--6, Sept 2013.

\end{thebibliography}

\end{document}